\documentclass[10pt,journal,compsoc]{IEEEtran}

\usepackage[OT1]{fontenc} 

\usepackage{booktabs}

\usepackage{wrapfig}
\usepackage{multirow}

\usepackage{makecell}

\usepackage{algorithm,algorithmic}
\usepackage{units}
\usepackage{graphicx}
\usepackage{amsmath}
\usepackage{color}

\begin{document}

\newcommand{\mydraft}{false}

\newcommand{\twofigure}[4]{
{\hbox{\includegraphics[width=#1,draft=\mydraft]{#3}\hspace{#2}\includegraphics[width=#1,draft=\mydraft]{#4}}}
}

\newcommand{\threefigure}[5]{
{\hbox{\hspace{-.045in}\includegraphics[width=#1,draft=\mydraft]{#3}\hspace{#2}\includegraphics[width=#1,draft=\mydraft]{#4}\hspace{#2}\includegraphics[width=#1,draft=\mydraft]{#5}}}
}

\newcommand{\fourfigure}[6]{
{\hbox{\includegraphics[width=#1,draft=\mydraft]{#3}\hspace{#2}\includegraphics[width=#1,draft=\mydraft]{#4}\hspace{#2}\includegraphics[width=#1,draft=\mydraft]{#5}\hspace{#2}\includegraphics[width=#1,draft=\mydraft]{#6}}}
}

\newcommand{\magicfigure}[3]{
{\hfill{\vspace{0.04cm}\hbox{\includegraphics[height=#2,draft=\mydraft]{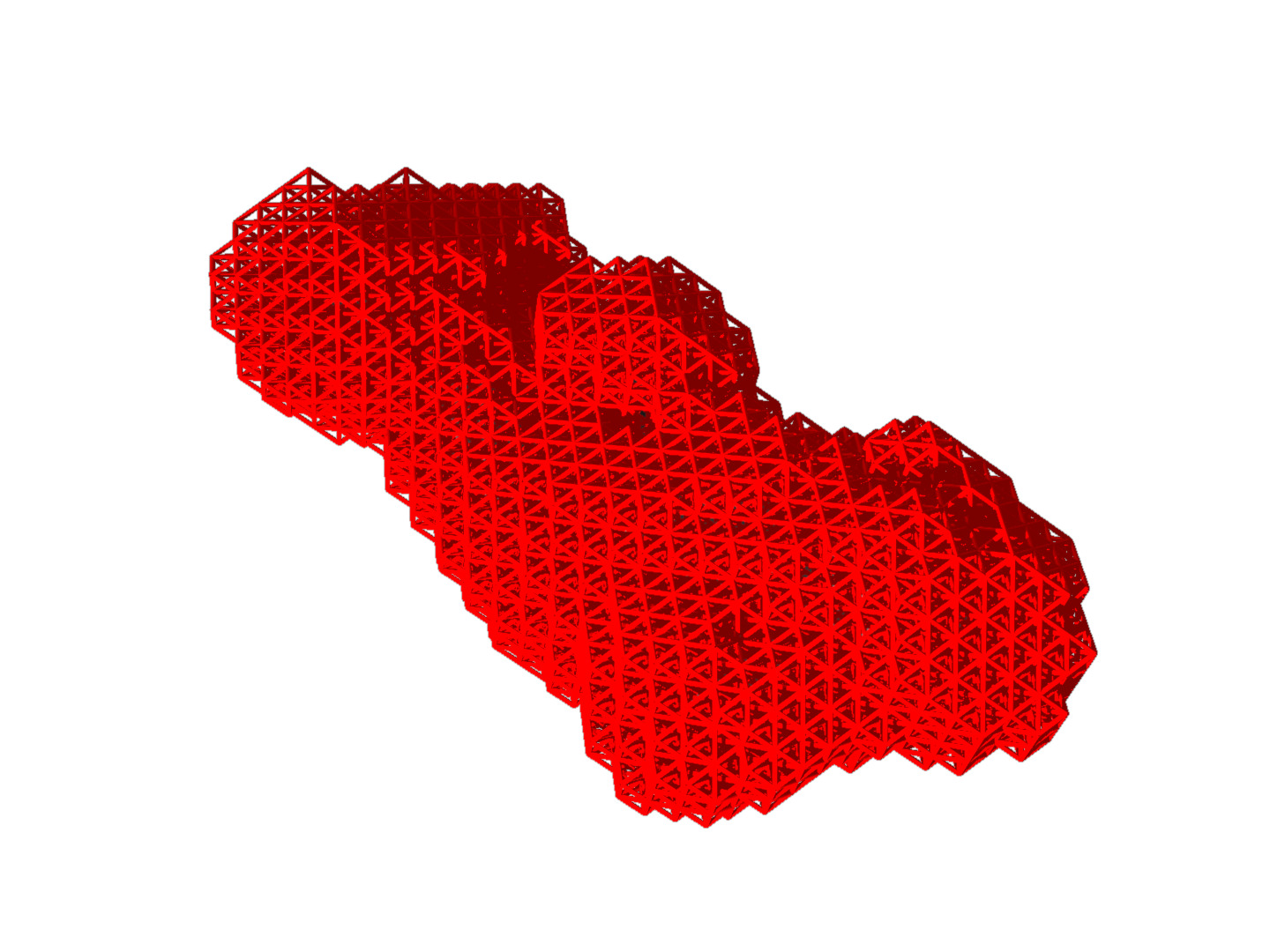}\hspace{#3}\vbox to #2{\hbox{\includegraphics[width=#1,draft=\mydraft]{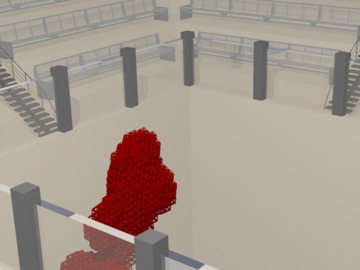}\hspace{#3}\includegraphics[width=#1,draft=\mydraft]{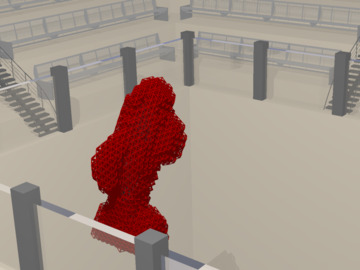}\hspace{#3}\includegraphics[width=#1,draft=\mydraft]{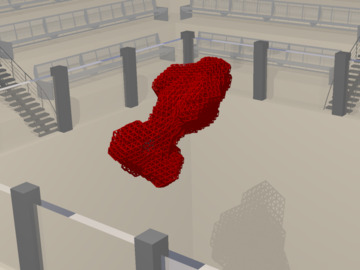}}\vfill\hbox{\includegraphics[width=#1,draft=\mydraft]{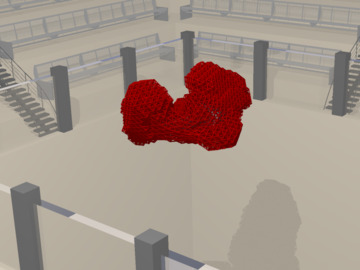}\hspace{#3}\includegraphics[width=#1,draft=\mydraft]{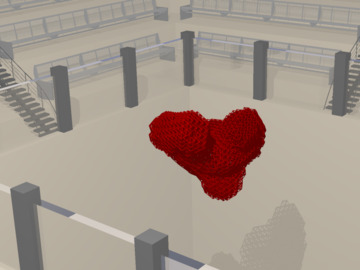}\hspace{#3}\includegraphics[width=#1,draft=\mydraft]{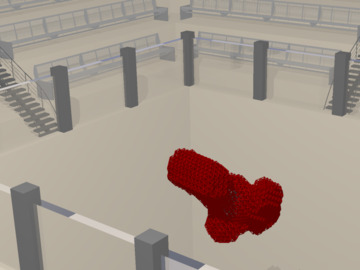}}}}}\hfill}
{\hfill\hbox{\includegraphics[height=#2,draft=\mydraft]{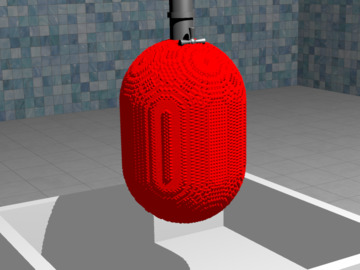}\hspace{#3}\vbox to #2{\hbox{\includegraphics[width=#1,draft=\mydraft]{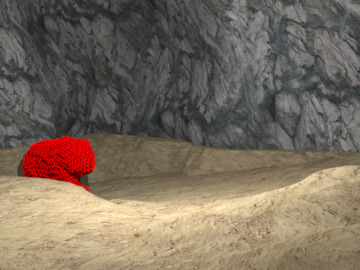}\hspace{#3}\includegraphics[width=#1,draft=\mydraft]{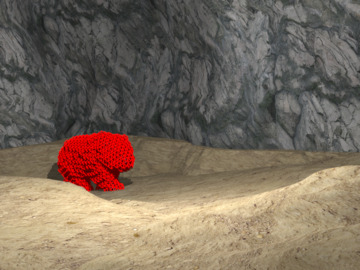}\hspace{#3}\includegraphics[width=#1,draft=\mydraft]{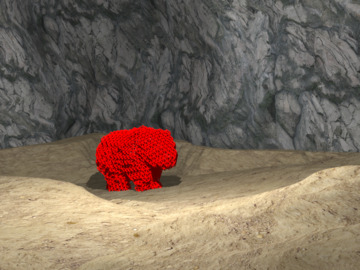}}\vfill\hbox{\includegraphics[width=#1,draft=\mydraft]{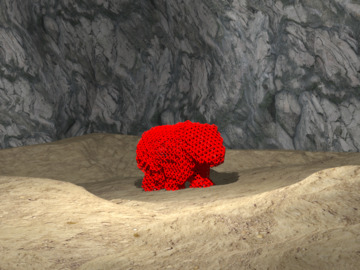}\hspace{#3}\includegraphics[width=#1,draft=\mydraft]{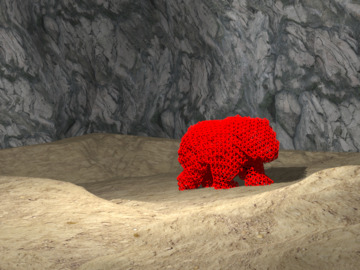}\hspace{#3}\includegraphics[width=#1,draft=\mydraft]{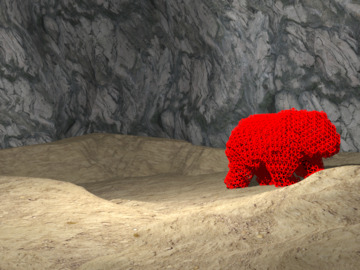}}}}\hfill}
}

\newif \ifdrawfigures
\drawfigurestrue

\newif \ifdisplayoutline
\displayoutlinefalse

\title{A Robust Volume Conserving Method for Character-Water Interaction} 

\author{Minjae Lee, 
David Hyde, 
Kevin Li, 
Ronald Fedkiw} 


\IEEEtitleabstractindextext{
\begin{abstract}
We propose a novel volume conserving framework for character-water interaction, using a novel volume-of-fluid solver on a skinned tetrahedral mesh, enabling the high degree of the spatial adaptivity in order to capture thin films and hair-water interactions.
For efficiency, the bulk of the fluid volume is simulated with a standard Eulerian solver which is two way coupled to our skinned arbitrary Lagrangian-Eulerian mesh using a fast, robust, and straightforward to implement partitioned approach.
This allows for a specialized and efficient treatment of the volume-of-fluid solver, since it is only required in a subset of the domain. The combination of conservation of fluid volume and a kinematically deforming skinned mesh allows us to robustly implement interesting effects such as adhesion, and anisotropic porosity.
We illustrate the efficacy of our method by simulating various water effects with solid objects and animated characters.
\end{abstract}

}

\maketitle

\IEEEdisplaynotcompsoctitleabstractindextext

\IEEEpeerreviewmaketitle


\section{Introduction}\label{sec:intro}

\begin{figure*}[t!]
    \centering
  \includegraphics[width=\linewidth]{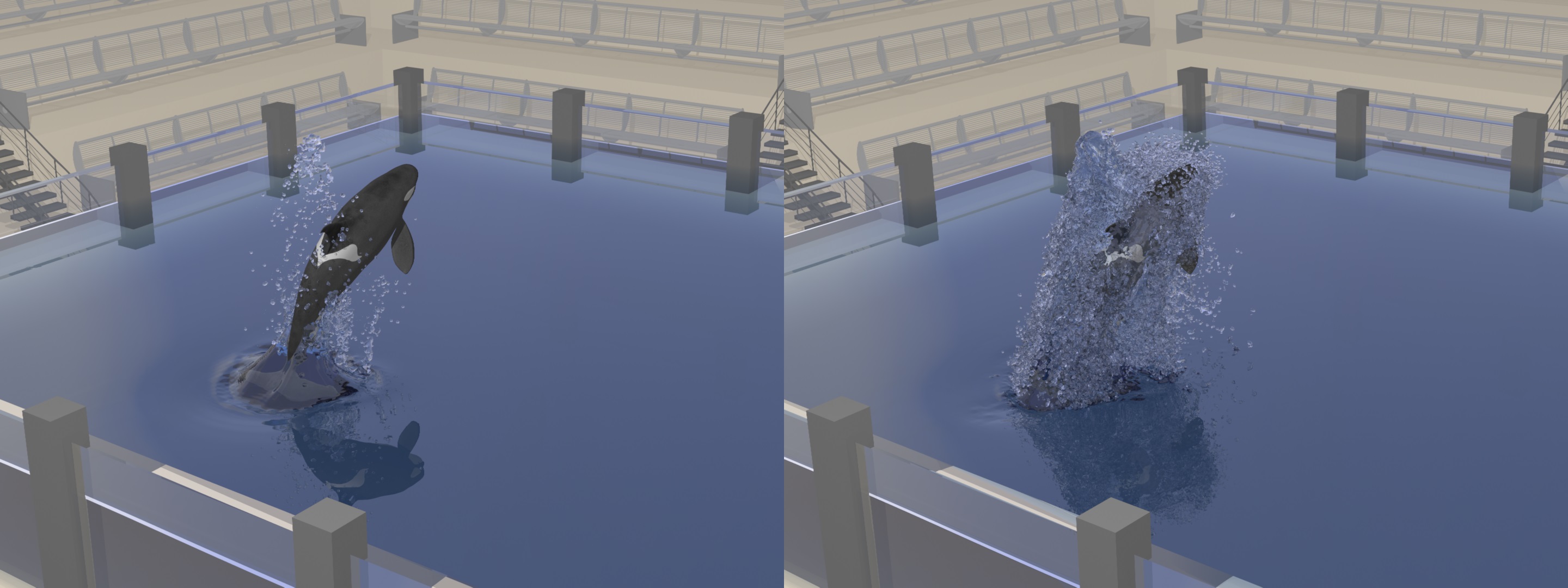}
  \centering
   \caption{(Left) Whale breaching with the PLS method using automatically generated removed particles for spray. Very little of the water volume follows the whale's motion because of volume loss on the relatively coarse Eulerian background grid. (Right) Using the same Eulerian grid, our ALE based VOF method on the KDSM produces much more visually interesting sheeting and spray effects.}
    \label{fig:teaserfigure}
\end{figure*}

\IEEEPARstart{C}{haracter-water} interaction is a widespread phenomenon in the visual effects industry, and there have been many efforts to push for higher quality water interaction with animated characters such as King Kong in \emph{Kong: Skull Island (2017)}, Hank the octopus in \emph{Finding Dory (2016)}, and various characters in \emph{Moana (2016)}.

Arguably, the most obvious approach to obtaining more detailed features anywhere in the domain is to place more degrees of freedom in the region of interest.
A number of adaptive methods have been developed such as Adaptive Mesh Refinement (AMR) \cite{Sussman:2003:ALS}, octree data structures \cite{losasso:2004:octree}, \cite{Losasso:2005:adaptive}, \cite{aan:2011:power}, lattice based tetrahedral methods \cite{Chentanez:2007:LSL}, \cite{Batty:2010:tetrahedral}, \cite{ando:2013:tetflip}, and Chimera grids \cite{english:2012:chimera}, \cite{english:2013:chimera_water}.
While these methods greatly improve water simulation detail through adaptivity, various authors have noted numerous drawbacks including the need to remesh very often, difficulties in implementation, performance bottlenecks induced by high communication costs, and issues related to domain decomposition due to a large number of small patches.
These issues are exacerbated when the adaptivity is required near boundaries with animated characters, since the character motion can rapidly change the region in space where the adaptivity is required.
A more natural approach would be to use an adaptive mesh that moves with the character such as the recently proposed kinematically deforming skinned mesh (KDSM) of \cite{lee:2018:kdsm_anon}.
This allows one to prebake the adaptivity so that on-the-fly refinement is not required during the simulation. This makes the method straightforward to implement and robust in its handling of delicate phenomena.

Even with additional degrees of freedom near the animated character, the highly dynamic water motion and thin films are notoriously difficult to simulate due in large part to both volume loss and difficulties with imposing proper boundary conditions between the water and the character.
We address volume conservation by proposing a novel volume-of-fluid (VOF) method implemented on the KDSM.
Although our proposed VOF method is novel, it is similar in spirit to other VOF methods such as \cite{mihalef:2004:breaking}, \cite{mihalef:2006:boiling} in that no fluid volume is lost, especially as compared to typical Eulerian methods.
VOF method is a well known technique as demonstrated in \cite{Hirt_Nichols:1981:VOF}, \cite{brackbill:1992:VOF}, \cite{rider:1998:volumetracking}, and \cite{Sussman:2000:CLSVOF}.
There have been some recent interesting works on boundary conditions between solids and fluids such as \cite{zhang:2016:resolving}, \cite{omar:2017:positive} using an Eulerian fluid grid (see \cite{Akinci:2013:VST} for SPH); however, it is more natural to specify these types of boundary conditions when the fluid grid is moving along with the solid in its Lagrangian frame, even if it is deforming a bit in that frame as is the case with KDSM.
With this treatment, much of the fluid moves along with the mesh being driven by the character animation (which is also driving the mesh) meaning that less fluid volume flows from one computational cell to another.
This is the typical arbitrary Lagrangian-Eulerian (ALE) approach, see for example \cite{feldman:2005:hybrid}, \cite{feldman:2005:deforming}, \cite{Klingner:2006:FAD}.
Notably, our method significantly differs from existing ALE implementations in that our ALE mesh is prebaked based on kinematically prescribed motion and has topology that remains consistent throughout the entire animation sequence.
This separation of the remeshing step resolves a key problem of ALE based methods which can lack robustness due to the numerical instabilities caused by ill-formed elements--this can now be addressed during a preprocessing step.

In order to increase the overall efficiency and efficacy of our approach, we only utilize the ALE based VOF method on the KDSM near the animated character while using a standard Eulerian based Cartesian grid solver in the rest of the domain, in our case the particle level set (PLS) method \cite{Enright:2002:PLS}, \cite{Enright:2002:AAR}, including spray particles \cite{losasso:2008:two}.
It is important to note that the PLS method is a hybrid method combining particle and grid representations, and early work was presented in \cite{Foster:1996:RAO} where they implemented a precursor to PIC/FLIP while removing unneeded interior particles far from the surface to boost performance and using level set to reconstruct smooth surface.
Recently, \cite{Ferstl:2016:NBF} proposed further improvements of PLS and FLIP hybrid method.
Thus, PLS and PIC/FLIP share important commonalities, and our VOF method can improve PLS or PIC/FLIP or any other method combining particle and grid representations.

Note that our spray particles carry mass and momentum as in \cite{losasso:2008:two} for visual quality when spray particles interacts with water surface instead of using massless particles as in vanilla PLS.
Importantly, the conservative nature of our VOF solver allows for relaxation of the numerical approach especially since the VOF solver is only required in a small subset of the domain near the character while a standard Eulerian solver is used elsewhere.
Thus, we devise a straightforward partitioned (as compared to monolithic) approach to the coupling of the fluid flow equations between the Eulerian Cartesian grid and the ALE based VOF solver on the KDSM; see Section \ref{sec:partitioned_coupling}.
Importantly, this combination of a standard Eulerian solver on the bulk of the domain with an ALE based KDSM mesh near the character allows the proposed VOF scheme to be incredibly simple, as discussed in Section \ref{sec:kdsm_fluids}, which is quite notable given the typical high level of complexity one usually confronts with VOF methods.

The first contribution of this paper is our strategy of prebaking the dense ALE mesh which occupies the space near the object or creature of interest, taking advantage of the adaptivity to capture detailed water phenomena based on the intuition that most interesting water effects are focused near the creature.
We achieve a robust simulation method by separating the nontrivial mesh processing operations from the simulation stage and incorporating them into a preprocessing stage, where we precompute various auxiliary data in order to improve the performance of the simulation.
Our second contribution is our novel VOF method, which conserves volume within the ALE mesh, whereas the PLS method in the background alone does not.
Our approach of conserving volume near the object or creature of interest allows us to implement various adhesion and porosity effects robustly and with mechanisms for artistic control.
The third contribution of our method is the straightforward partitioned approach for coupling the coarse background Eulerian grid and our fine ALE mesh, which greatly streamlines the development process.

\begin{figure*}[t!]
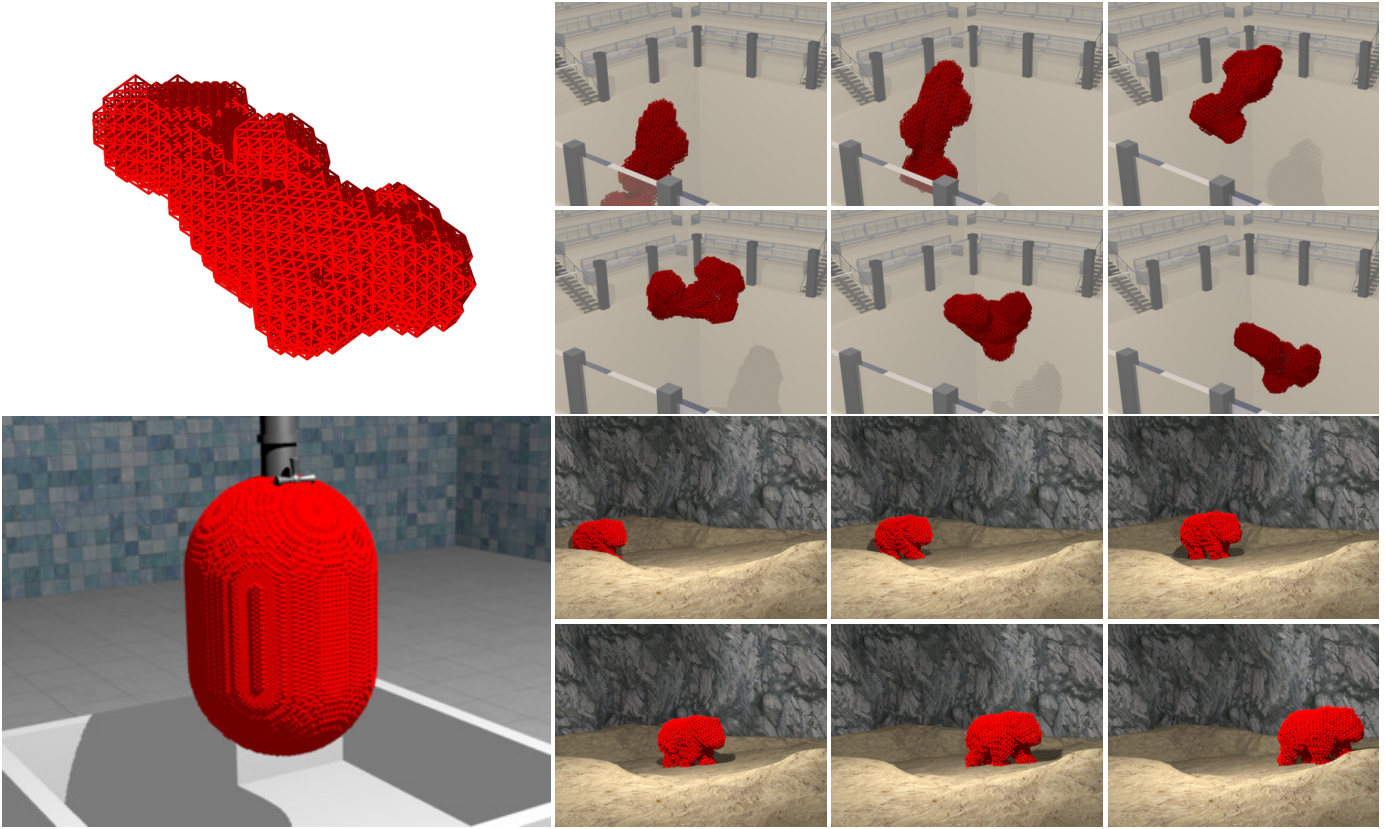

  \centering
  \magicfigure{.1975\textwidth}{.3\textwidth}{.03in}
  \caption{(Top Left) A KDSM mesh around a whale in a normalized pose (also known as T-pose or rest pose). (Top Right) A sample animation showing the KDSM skinned to follow an animation of a whale breaching. (Bottom Left) A KDSM mesh around the ball. (Bottom Right) A sample animation showing the KDSM skinned to follow an animation of a bear walking on a shore.}
  \label{fig:whale_kdsm}
\end{figure*}

\section{KDSM} \label{sec:kdsm}

Following \cite{lee:2018:kdsm_anon}, we generate a KDSM from a tetrahedral BCC (body centered cubic) lattice as in \cite{molino:2003:meshing} using a thickened level set of the triangulated surface skin mesh of a creature or an object in a normalized pose (see Figure \ref{fig:whale_kdsm} left).
Then, given an animation sequence of the creature's triangulated surface skin mesh, the KDSM nodes inside the creature are morphed to follow the animation as per \cite{ali-hamadi:2013:anatomy,cong:2015:fully} capturing the kinematic deformation of the creature's skin and its volumetric interior.
We connect the KDSM nodes that are exterior to the skin mesh of the creature to one another and to the internal nodes via a constitutive model (e.g.\ mass spring), so that the KDSM nodes that are external to the creature also follow the animation (see Figure \ref{fig:whale_kdsm} right).
In addition, zero length spring attachments are connected between the creature's skin mesh and corresponding barycentric locations in the KDSM in order to obtain more accurate deformations of the KDSM near the creature's skin.

\cite{lee:2018:kdsm_anon} embedded hair particles in the KDSM so that they would follow the skinned animation sequence.
Each hair's base particle is embedded on the surface of the character's triangle mesh skin and the rest of hair particles are embedded in the tetrahedral mesh using hard bindings as in \cite{sifakis:2007:hybridsolids}.
They also showed how a duplicated KDSM which is following the original constrained with zero length springs could be used to produce more dynamic hair behavior.
Since the original KDSM sequence contains a significant portion of the desired motion, further iterations for dynamic hair behavior becomes much more efficient.
In addition, they showed the benefits of the KDSM when simulating individual hairs as well as how to use the KDSM to implement a blendshape system including the effects of clumping, sagging, and matting.
They simulated individual hairs using \cite{selle:2008:hair} and soft bindings (see \cite{sifakis:2007:hybridsolids}) connecting hair particles and their corresponding desired positions in the KDSM via zero length springs.
They proposed a shape-preserving tetrahedral column to maintain the original style of the hair also connected to hair particles via zero length springs.
The blendshape hair component is implemented as a collection of barycentric coordinates for each hair particle, and can be interpolated in time to implement a change in hairstyle over time (e.g. bear getting wet in the water).
Then length preservation and interpenetration is run as a post-process, shortening the each hair segment to its rest length (see \cite{muller:2012:fast,sanchez:2015:real}) and applying pushout method from \cite{bridson:2003:cloth}.
Interestingly, they demonstrated how the air volume enclosed by the KDSM could be utilized to add adhesion and drag effects, porosity for hair, etc.

\begin{figure}[t]
    \centering
    \hbox{
        \includegraphics[width=0.235\textwidth,draft=\mydraft]{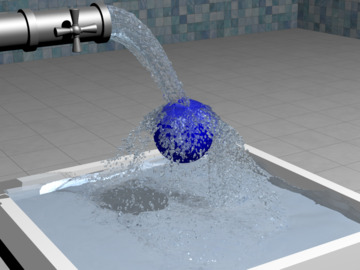}
        \includegraphics[width=0.235\textwidth,draft=\mydraft]{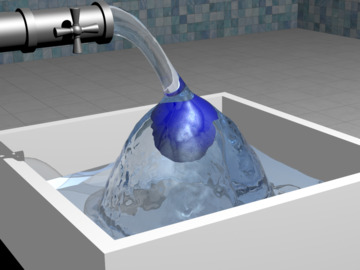}
    }
    \vspace{.01in}
    \hbox{
        \includegraphics[width=0.235\textwidth,draft=\mydraft]{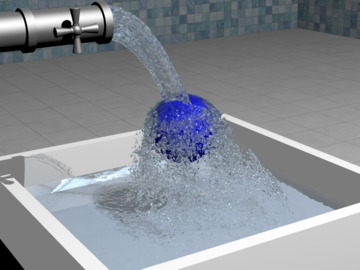}
        \includegraphics[width=0.235\textwidth,draft=\mydraft]{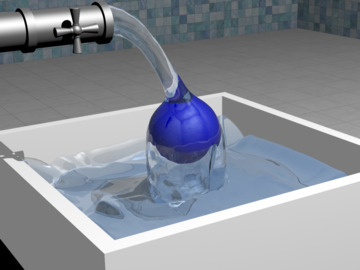}
    }
    \caption{
        (Top Left) A standard PLS simulation on a relatively coarse Eulerian grid. (Top Right) Our ALE based VOF method on the KDSM achieves better water sheeting and volume conservation using the same Eulerian grid.
        (Bottom Row) Applying adhesion forces to both simulations produces the desired clinging to the ball with our method but has almost no effect on the PLS simulation.
    }
    \label{fig:adhesion}
\end{figure}

A major shortcoming of this prior work in regards to water simulation is that the KDSM is merely used to provide information such as forces that augment the treatment of the Eulerian grid.
Thus, all the typical drawbacks of volume loss, etc., are not only still present but potentially worsened by these additional forces.
See, for example, Figure \ref{fig:adhesion} and \ref{fig:adhesion_top}.

\section{KDSM Fluids} \label{sec:kdsm_fluids}

Our novel ALE based VOF method on the KDSM produces compelling results even though it has none of the complexities associated with typical VOF methods--even the volume conservation step is quite simple.
We stress that the ability to use such a simple method is due in large part to the partitioned coupling discussed in Section \ref{sec:partitioned_coupling}, the adaptivity of the KDSM, exact volume conservation, and the Lagrangian nature of the KDSM as it follows the animated creature.
Our VOF method fully conserves volume within the KDSM, while the rest of the domain simulated using the PLS method does not.

\subsection{Precomputation} \label{sec:precomputation}

Starting from the original KDSM animation, we precompute auxiliary information such as adaptivity by subdividing the KDSM until a desired resolution is reached (for our examples, we subdivided KDSM multiple times until the size of tetrahedron is 4 to 8 times smaller than the background grid cell size).
Prebaking subdivisions to obtain per-frame ALE meshes with consistent topology greatly increases the robustness and minimizes computation time as compared to typical ALE remeshing.
We subdivide using \cite{molino:2003:meshing}, but only utilized the subdivision operation part without any adaptivity features as we wanted an evenly subdivided ALE mesh.
Although we opt to precompute our KDSM, one could subdivide on-the-fly if desired.
Note that we use a relatively coarse KDSM to run a mass spring simulation and then subdivide the resulting KDSM for performance reasons.
One can always use a dense KDSM to obtain better boundaries near the creature's skin, but we found the current scheme sufficient for our examples.
We additionally precompute a number of useful quantities such as per node velocities and a level set volume for every frame of the animation.
Every cut cell tetrahedron, those containing a part of the creature's surface, is assigned a surface normal and object velocity.
Those surface normals and object velocities are extrapolated to every tetrahedron of the KDSM exterior to the creature using the level set information.

Per tetrahedron solid occupancy is precomputed in the cut cells using point samples and the quadrature formula of \cite{zhang:2009:quadrature} testing how many point samples are inside of the creature using the level set representation.
Instead of using the exact volume, we simply use the fraction of point samples inside and outside the creature to compute an approximate volume.
This added simplicity is equivalent to a slight sub-tetrahedral perturbation of the solid surface.
Obviously, this could be done more accurately, but we found that this simple method worked quite well, and simplicity and efficiency is highly desirable when one might want to refine near the surface of the creature on-the-fly.

Our proposed volume conservation method is motivated by the shock propagation for rigid bodies from \cite{guendelman:2003:rigid}.
As such, we precompute the rank of each tetrahedron as its topological distance from the creature, similar in spirit to the contact graph from \cite{guendelman:2003:rigid}.
Tetrahedra fully inside the creature are assigned a rank of $-1$, and partially filled tetrahedra are assigned rank $0$.
Then, all tetrahedra with unassigned ranks that are node neighbors of rank $0$ tetrahedra are assigned rank $1$.
Rank $2$, rank $3$, etc. are assigned similarly.

\begin{figure}
    \centering
    \hbox{
        \includegraphics[width=0.235\textwidth,draft=\mydraft]{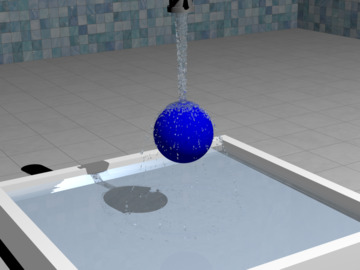}
        \includegraphics[width=0.235\textwidth,draft=\mydraft]{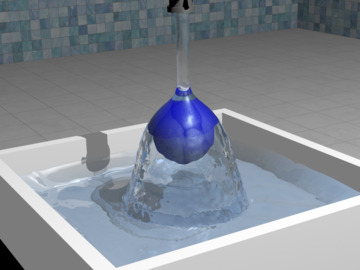}
    }
    \vspace{.01in}
    \hbox{
        \includegraphics[width=0.235\textwidth,draft=\mydraft]{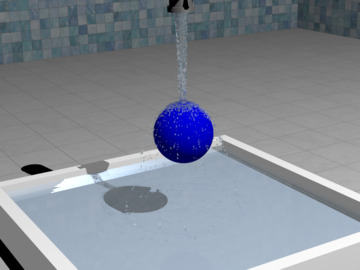}
        \includegraphics[width=0.235\textwidth,draft=\mydraft]{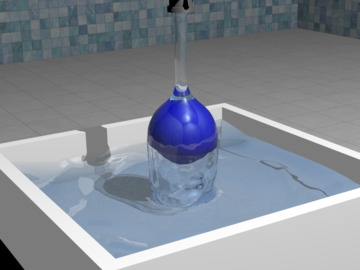}
    }
    \caption{
        Same as Figure \ref{fig:adhesion}, but using an even smaller water stream accentuating the benefits of our approach especially when considering volume conservation.
    }
    \label{fig:adhesion_top}
\end{figure}

\subsection{Advection} \label{sec:advection}

We store vector valued velocities as per tetrahedron values, and multiplying by the mass of water in a tetrahedron yields momentum.
Our ALE based VOF method updates the velocity and the amount of water in each tetrahedron on a new mesh given values on an old mesh.

\begin{figure}
    \centering
    \includegraphics[width=0.235\textwidth,draft=\mydraft]{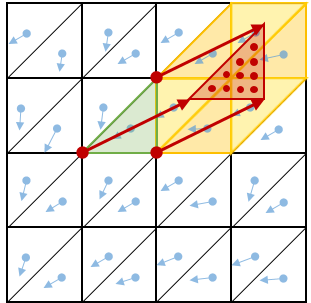}
    \caption
    {
        A backtraced triangle shown in red uses each of its 10 quadrature point samples to transport water from the old mesh to the new one.
        Here, we render both the old and the new mesh in the same location assuming that the KDSM is not moving for simplicity of depiction.
        Each red dot point sample would attempt to remove 1/10 of the area of the red triangle from the specific yellow triangle that it is interior to.
    }
    \label{fig:advection}
\end{figure}

First, for each tetrahedron, we trace its nodes backwards in time using its vector valued velocity multiplied by $\Delta t$.
As shown in Figure \ref{fig:advection}, this rigidly translates the tetrahedron.
Alternatively, one could instead compute per node velocities, but we have found this unnecessary.
If a backtraced node collides with a solid surface, it is clamped to that location.
Thus, the backtraced tetrahedron does not deform unless it hits a solid surface.
The resulting backtraced tetrahedron (shown in red in Figure \ref{fig:advection}) is used to collect water from the old mesh in order to deposit it on the new mesh.
Instead of performing the usual complex geometric intersections between backtraced tetrahedra and the old mesh, we take a simpler approach using a number of quadrature formula point samples (again from \cite{zhang:2009:quadrature}).
Each point sample (shown as a red dot in Figure \ref{fig:advection}) attempts to transport a certain amount of water from the old mesh tetrahedron it lies within (shown in yellow in Figure \ref{fig:advection}) to the original tetrahedron on the new mesh (shown in green in Figure \ref{fig:advection}).
This potential amount of transported water is calculated as the volume of the backtraced tetrahedron divided by its number of point samples.
Both water volume and associated momentum are transported.
If a point sample falls outside the KDSM, we compute the amount and momentum of water to transport using interpolation from the background Eulerian grid.
Note that this water is not removed from the background grid, since the background grid is treated in an Eulerian fashion and updated properly via the coupling proposed in Section \ref{sec:partitioned_coupling}.

The aforementioned advection might attempt to transport more water out of a tetrahedron on the old mesh than that tetrahedron contains.
This could occur because of size differences between tetrahedra or because multiple point samples from separate tetrahedra of the new mesh request water from the same tetrahedron on the old mesh.
Thus, in a second step, we visit every tetrahedron on the old mesh and scale down the amount of water each point sample removes in order to match the total volume of water in this old mesh tetrahedron as in \cite{Lentine:2011:MMCFS}.

In a third step, we identify any tetrahedra on the old mesh that may have excess water, which was not transported to the new mesh, and forward advect this water to the new mesh similar to \cite{Lentine:2011:MMCFS}, \cite{lentine:2012:water}, \cite{Klingner:2006:FAD}.
However, our method can be much simpler because we track the exact volume of water with our VOF method, and therefore do not mind overfilling tetrahedra because they are drained to their appropriate volume during the volume conservation step (see Section \ref{sec:projection}).
For each tetrahedron on the old mesh that requires forward advection, we use that tetrahedron's velocity to advect its nodes forward in time (in the opposite direction of Figure \ref{fig:advection}) and use its point samples to locate which tetrahedra in the new mesh will receive its water.
Similar to backward advection, we clamp nodes that collide with solids, use the number of point samples and the original tetrahedron volume to decide on the fraction of water deposited in each target tetrahedron, and utilize special treatment for any point sample that leaves the KDSM and lands on the background Eulerian grid.
In particular, we find particle creation to be quite useful for transporting water off of the KDSM (see Section \ref{sec:particle_generation}).

\subsection{Volume Conservation} \label{sec:projection}

Our volume conservation method is used to enforce incompressibility on the new mesh using our precomputed rank.
The method consists of three parts: smear, pushout, and velocity correction.
Pushout transports excess water and associated momentum outward from the creature's skin mesh using rank motivated by \cite{guendelman:2003:rigid}.
However, this ignores the fact that the fluid can rotate and move laterally, so we first apply what we refer to as a smearing step to account for this behavior.
Both the smear and the pushout step transport the volume and associated momentum together.
Finally, velocity correction is used to apply boundary conditions on the water from the creature.

\begin{figure}
    \centering
    \includegraphics[width=0.235\textwidth,draft=\mydraft]{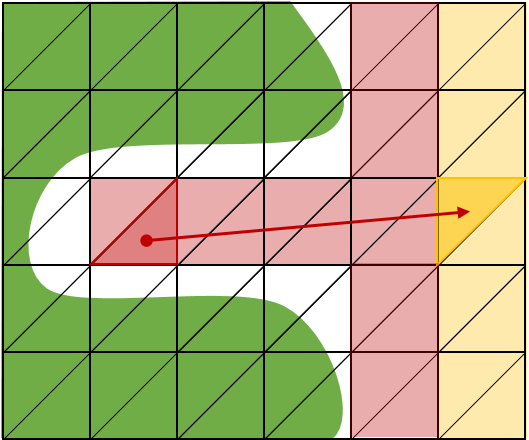}
    \caption{
        The triangles cut by the green solid surface region would be assigned rank 0, their node neighbors shown in red would be rank 1, and their node neighbors shown in yellow would be rank 2.
        The arrow shows how a rank 1 triangle needs to look non-locally in order to find a rank 2 triangle that it can deposit excess water into.
    }
    \label{fig:projection}
\end{figure}

We refer to tetrahedra as oversaturated when they contain more water than their volume should allow.
The smearing step loops over oversaturated tetrahedra, except those on the boundary of the KDSM, and distributes the excess fluid equally to face neighbors.
The boundary tetrahedra are taken care of in the pushout phase.

For pushout, we iterate over oversaturated tetrahedra in order from the lowest rank to the highest starting with tetrahedra of rank $0$ which intersect the creature's skin mesh.
For each oversaturated tetrahedron, we distribute as much excess fluid as possible equally to its face neighbors that are not yet fully saturated.
If there is still excess fluid after every neighbor is saturated, it is distributed equally to all face neighbors with strictly higher ranks.
Note that it is possible to have a tetrahedron without any valid neighbors to distribute water to, since the creature skin mesh can have narrow space between solid surfaces as illustrated in Figure \ref{fig:projection}.
To handle this case, we preprocess the neighbor information using breadth first search to look for non-local neighbors with higher ranks to properly transport excess water to.
Although not trivial, this can be done in the preprocessing step after determining rank.
For boundary tetrahedra, we transfer excess fluid to the background Eulerian grid for particles as discussed in Section \ref{sec:partitioned_coupling} and \ref{sec:particle_generation}, respectively.

Finally, velocity correction is used to apply boundary conditions on the fluid from the creature.
We assign a Boolean flag per tetrahedron indicating whether its fluid velocity needs to be corrected, and initialize all flags to false.
Then, we iterate over all cut cell tetrahedra with water and set flags to true.
Subsequently, we loop over the tetrahedra in the same order as in the pushout phase, clamping the normal component of the fluid velocity to be the precomputed object normal velocity when the flag is set to true and the normal component of velocity is smaller than the precomputed object normal velocity.
Higher rank tetrahedra have their Boolean flag set to be true when they have a lower rank face neighbor which is fully saturated that required clamping.
This limits the enforcement of boundary conditions to those tetrahedra exposed to the creature surface by a column of water.
Although this removes circulation on the KDSM, the circulation is restored from the background Eulerian grid during coupling as discussed in Section \ref{sec:partitioned_coupling}.

\begin{figure}
    \centering
    \hbox{
        \includegraphics[width=0.156\textwidth,draft=\mydraft]{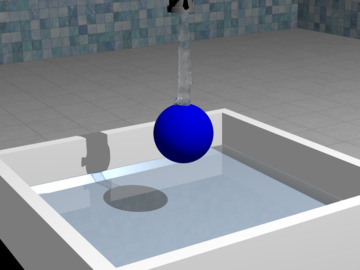}
        \includegraphics[width=0.156\textwidth,draft=\mydraft]{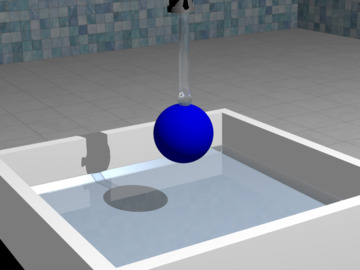}
        \includegraphics[width=0.156\textwidth,draft=\mydraft]{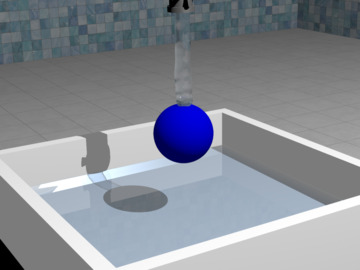}
    }
    \vspace{.01in}
    \hbox{
        \includegraphics[width=0.156\textwidth,draft=\mydraft]{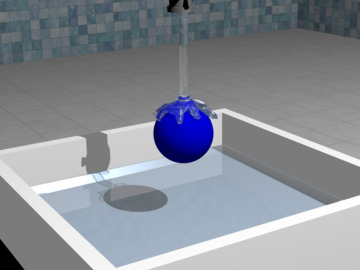}
        \includegraphics[width=0.156\textwidth,draft=\mydraft]{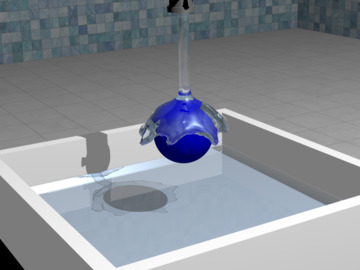}
        \includegraphics[width=0.156\textwidth,draft=\mydraft]{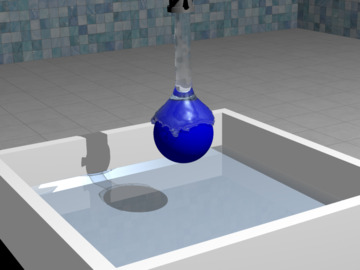}
    }
    \vspace{.01in}
    \hbox{
        \includegraphics[width=0.156\textwidth,draft=\mydraft]{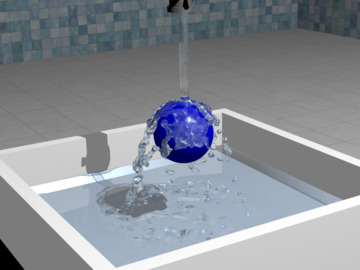}
        \includegraphics[width=0.156\textwidth,draft=\mydraft]{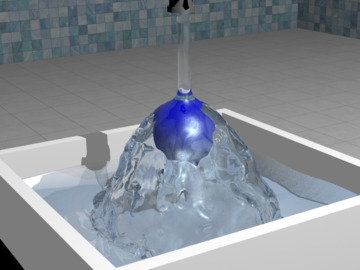}
        \includegraphics[width=0.156\textwidth,draft=\mydraft]{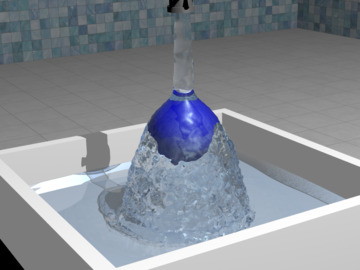}
    }
    \caption{
        (Left Column) Our VOF method with a naive projection implementation which does not conserve volume.
        (Middle Column) Our VOF method with smear and pushout while replacing our velocity correction step with a standard Poisson solver.
        (Right Column) Our VOF method with proposed smear, pushout, and velocity correction steps. The middle and right columns conserve volume.
    }
    \label{fig:pressure}
\end{figure}

This approach is not a standard projection scheme, but still enforces a divergence free condition thereby enforcing incompressiblity. One other notable approach that is not an advection-projection scheme is \cite{Zehnder:2018:ASD}.
To elaborate, fluid simulated using the Navier-Stokes equations is assumed to obey the conservation of mass equation $\partial \rho / \partial t + \nabla \cdot (\rho \mathbf{u}) = 0$ (i.e. no fluid is created or destroyed).
Here, $\mathbf{u}$ is the fluid velocity, $t$ is time, and $\rho$ is the fluid density.
By the product rule, this is equivalent to $\partial \rho / \partial t + \rho \nabla \cdot \mathbf{u} + \mathbf{u} \cdot \nabla \rho = 0$.
Using this equation, it can be seen that setting $\nabla \cdot \mathbf{u} = 0$ is equivalent to setting $\partial \rho / \partial t + \mathbf{u} \cdot \nabla \rho = 0$.
Either condition implies the other--they are equivalent from the conservation of mass.
Setting $\rho = m / V$ and using the product rule gives $(1 / V)(\partial m / \partial t) - (m / V^2)(\partial V / \partial t) + \mathbf{u} \cdot ((1 / V)\nabla m - (m / V^2)\nabla V) = 0$, which can be regrouped as $(1 / V)(\partial m / \partial t + \mathbf{u} \cdot \nabla m) - (m / V^2)(\partial V / \partial t + \mathbf{u} \cdot \nabla V) = 0$.
The first term $(1 / V)(\partial m / \partial t + \mathbf{u} \cdot \nabla m)$ must equal zero since mass is conserved along streamlines.
This means $\partial V / \partial t + \mathbf{u} \cdot \nabla V$ must also equal zero, and taking this with the divergence free condition $\nabla \cdot \mathbf{u} = 0$ yields $\partial V / \partial t + \nabla \cdot (\mathbf{u} V) = 0$, i.e. conservation of volume.
Thus, conserving volume is equivalent to enforcing a divergence-free velocity field.

With regard to maintaining the incompressibility of a simulated fluid, we note that a standard projection scheme such as the classic method introduced by \cite{chorin:1968:NNS} is just one proposed algorithm for this task.
In fact, Chorin advocated at least two distinct schemes \cite{chorin:1967:ANM}, though the advection-projection procedure became most popular.
Chorin-style projection claims that one can advect a fluid state ad-hoc off of the manifold of all incompressible fluid fields and then correct the ensuing error by projecting back onto that manifold.
However, these projection-style schemes are known to be brittle (e.g. they require small time steps and do not even always converge under temporal refinement), and they are well-known to be unable to capture important physics of fluids (such as viscosity, due to its parabolic nature).
Thus, solving a pressure Poisson equation to enforce a divergence-free velocity field is not the ultimate, and certainly not the only, numerical scheme to simulate incompressible flow.
Our technique, which indeed differs from a pressure projection, is able to successfully and robustly conserve volume, which as shown above implies the standard incompressibility condition.
Hence our volume conservation method maintains a faithful relationship with the underlying fluid principles and equations.
Moreover, at a high level, we remark that even the Navier-Stokes equations quickly fail to be a physically accurate model when considering real-world flow problems i.e. turbulence; however, such approximations are heavily relied upon in computer graphics due to their computational amenability and their ability to produce visually plausible results.

\begin{figure}
    \centering
    \hbox{
        \includegraphics[width=0.235\textwidth,draft=\mydraft]{figures/ball_top_vof_adhesion.jpg}
        \includegraphics[width=0.235\textwidth,draft=\mydraft]{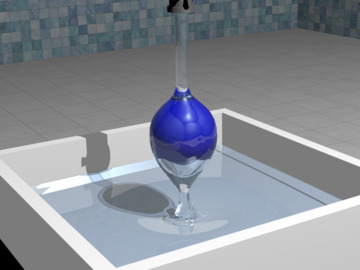}
    }
    \vspace{.01in}
    \hbox{
        \includegraphics[width=0.235\textwidth,draft=\mydraft]{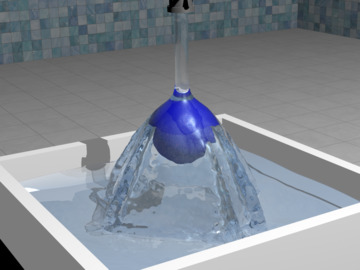}
        \includegraphics[width=0.235\textwidth,draft=\mydraft]{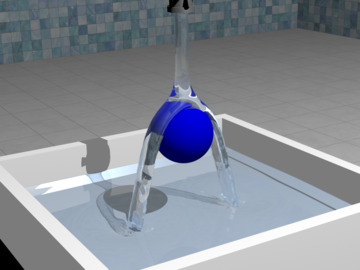}
    }
    \caption{
        Our ALE based VOF method provides robust adhesion control to produce various effects.
        All four of these examples are obtained by merely varying adhesion effects.
    }
    \label{fig:adhesion_all}
\end{figure}

In order to evaluate our method compared to other approaches and to explore possible extensions, we implemented a standard Poisson solver by assigning pressures on nodes similar to \cite{ando:2013:tetflip}.
This implementation solves the inviscid, incompressible Navier-Stokes equations, $\partial \mathbf{u} / \partial t = -(\mathbf{u} \cdot \nabla) \mathbf{u} -\nabla p / \rho + \mathbf{f}$ while satisfying $\nabla \cdot \mathbf{u} = 0$ to enforce the divergence free condition for the velocity field without any advanced modifications ($p$ is pressure, $\mathbf{f}$ is external forces).
We ran two different flavors of this alternative; one is to completely replace our volume conservation scheme with the standard Poisson solver ignoring the volume conservation entirely within the projection, and the other is to replace only the velocity correction while keeping smear and pushout to conserve volume.
Note that the smear and pushout steps transport fluid with its momentum, so oversaturated fluid velocity propagates to its neighbors.
Thus, the second version spreads water outward more than the first version.
We ran all implementations on the KDSM with the same setup, and the results are shown in Figure \ref{fig:pressure}.
In Figure \ref{fig:pressure}, we found that the right column is more desirable than the left because it conserves volume, and is faster and more robust than the middle column since we do not have to solve a linear system.

\subsection{Adhesion} \label{sec:adhesion}

We allow an artist to paint adhesion coefficients $\alpha$ and force directions $\vec{d}$ on the triangulated surface mesh of the creature, and then we rasterize this information to the KDSM setting adhesion quantities in rank $0$ tetrahedra.
We propagate adhesion quantities to face neighbors (or non-local neighbors) of strictly higher rank tetrahedra by averaging the adhesion quantities from lower rank neighbors for which adhesion had already been specified.
We apply an adhesion force $\alpha \vec{d}$ when a tetrahedron is within a prescribed distance $\phi_a$ from the creature's surface with linear falloff, i.e.\ $\alpha (\phi_a - \phi)/\phi_a \vec{d}$ where $\phi$ is the distance from the creature's surface (similar to \cite{zhu:2014:codimensional}).

\begin{figure*}
    \centering
    \hbox{
        \includegraphics[width=0.245\textwidth,draft=\mydraft]{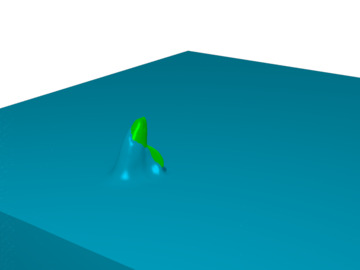}
        \includegraphics[width=0.245\textwidth,draft=\mydraft]{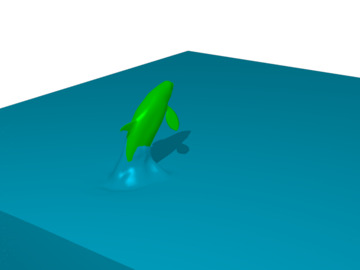}
        \includegraphics[width=0.245\textwidth,draft=\mydraft]{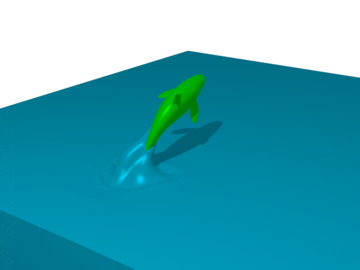}
        \includegraphics[width=0.245\textwidth,draft=\mydraft]{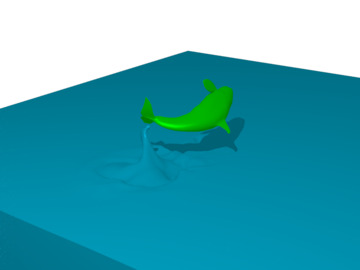}
    }
    \vspace{.017in}
    \hbox{
        \includegraphics[width=0.245\textwidth,draft=\mydraft]{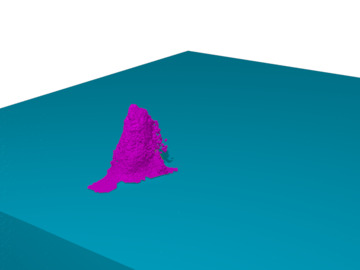}
        \includegraphics[width=0.245\textwidth,draft=\mydraft]{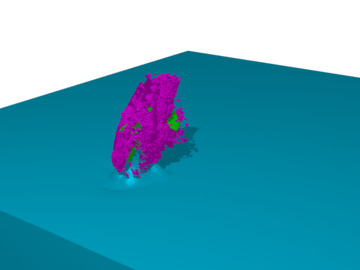}
        \includegraphics[width=0.245\textwidth,draft=\mydraft]{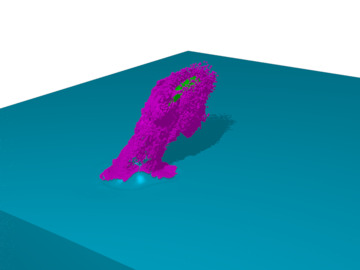}
        \includegraphics[width=0.245\textwidth,draft=\mydraft]{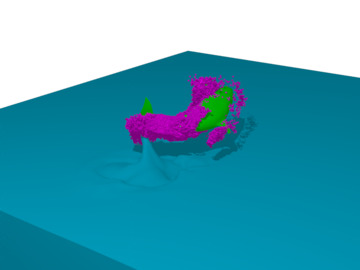}
    }
    \vspace{.017in}
    \hbox{
        \includegraphics[width=0.245\textwidth,draft=\mydraft]{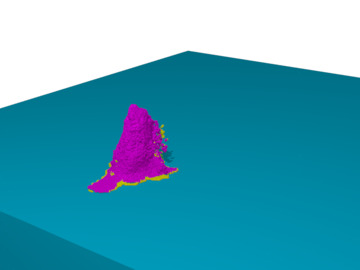}
        \includegraphics[width=0.245\textwidth,draft=\mydraft]{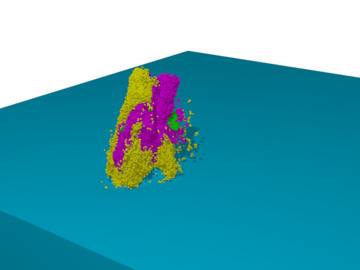}
        \includegraphics[width=0.245\textwidth,draft=\mydraft]{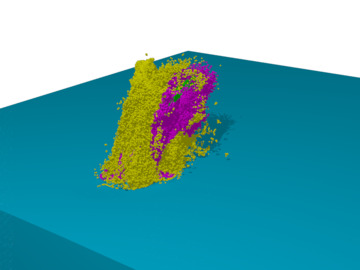}
        \includegraphics[width=0.245\textwidth,draft=\mydraft]{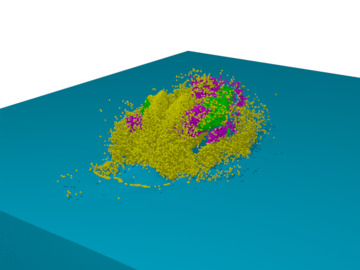}
    }
    \vspace{.017in}
    \hbox{
        \includegraphics[width=0.245\textwidth,draft=\mydraft]{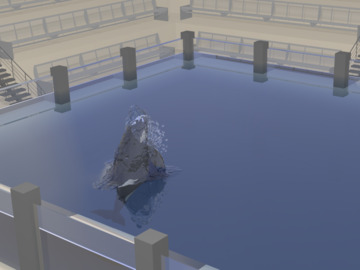}
        \includegraphics[width=0.245\textwidth,draft=\mydraft]{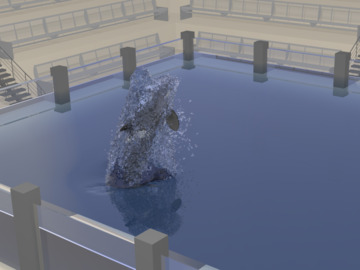}
        \includegraphics[width=0.245\textwidth,draft=\mydraft]{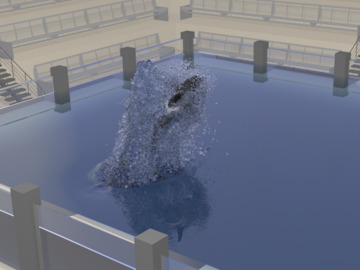}
        \includegraphics[width=0.245\textwidth,draft=\mydraft]{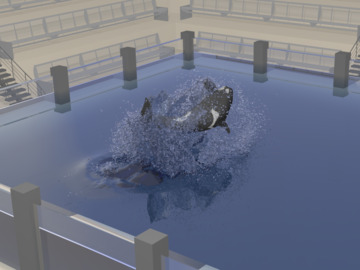}
    }
    \caption{
        A whale breaches out of the water.
        (Top Row) Visualization of the Eulerian water.
        (Second Row) VOF tetrahedra water is rendered in pink.
        (Third Row) Particles are shown in yellow.
        (Fourth Row) Final rendering.
    }
    \label{fig:whale}
\end{figure*}

Figure \ref{fig:adhesion_all} illustrates some of the many interesting visual effects obtainable by varying adhesion parameters.
Notably, it is the robust volume conservation of our VOF method and the adaptivity of the KDSM that allows for such interesting effects.
We attempted similar simulations using a standard Eulerian method and mostly achieved disturbing volume loss.
The top row shows how increasing adhesion (from left to right) makes the water to stick to the ball and flow around to the bottom surface before separating.
The bottom left image was created with vectors $\vec{d}$ pointing outwards from the ball at various locations to produce thickened streams.
In contrast, the bottom right figure shows how the vectors $\vec{d}$ can be used to direct water away from parts of the ball's surface, drying it out.

\section{Partitioned Coupling} \label{sec:partitioned_coupling}

We utilize three different representations for water: besides the VOF representation on the KDSM, we also use both free particles and velocities on the background Eulerian Cartesian grid as is typical for the standard PLS method (see e.g.\ \cite{Enright:2002:AAR}).
See Figure \ref{fig:whale}.
Our partitioned coupling method consists of four major steps.
In the first step, each of our three representations (VOF tetrahedra, particles, and Eulerian Cartesian grid) are advected forward in time.
The method of Section \ref{sec:advection} is used for the VOF tetrahedra, while the standard PLS method is used to advect the Eulerian grid velocities and to move the particles.
In a second step, momentum is transferred between the three representations in order to maximize the visual efficacy of the results.
Then, external forces are independently added to each representation, before projecting the velocity into a divergence free state acceptable to all three representations.
The steps are summarized below:

\begin{enumerate}
  \item Advection (each stage is independent)
  \begin{enumerate}
    \item VOF advection (Section \ref{sec:advection})
    \item Particle advection
    \item Eulerian advection
  \end{enumerate}
  \item Momentum transfer
  \begin{enumerate}
    \item Transfer momentum from VOF to Eulerian (optional)
    \item Transfer momentum from Eulerian to VOF
    \item VOF particle reincorporation
  \end{enumerate}
  \item Add external forces (each stage is independent)
  \begin{enumerate}
    \item VOF external forces
    \item Particle external forces
    \item Eulerian external forces
  \end{enumerate}
  \item Volume conservation
  \begin{enumerate}
    \item VOF volume conservation (Section \ref{sec:projection})
    \item Eulerian projection
    \begin{enumerate}
      \item Eulerian particle reincorporation
      \item Projection
    \end{enumerate}
    \item Transfer momentum from Eulerian to VOF
  \end{enumerate}
\end{enumerate}

\begin{figure}[b]
    \centering
    \hbox{
        \includegraphics[width=0.235\textwidth,draft=\mydraft]{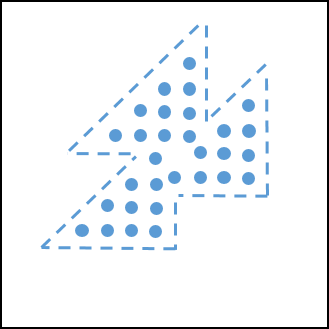}
        \includegraphics[width=0.235\textwidth,draft=\mydraft]{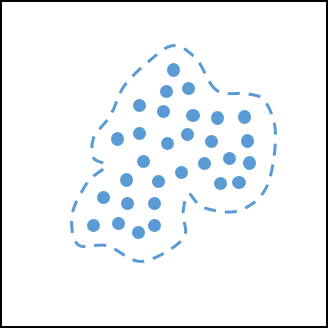}
    }
    \vspace{.01in}
    \hbox{
        \includegraphics[width=0.235\textwidth,draft=\mydraft]{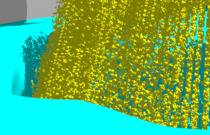}
        \includegraphics[width=0.235\textwidth,draft=\mydraft]{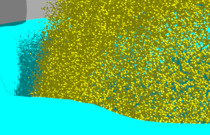}
    }
    \caption{
        Top row compares the particle positions obtained with uniform versus jittered sampling emphasizing how well our eyes capture structured information (even when we do not want them to).
        Bottom row compares the two approaches for an actual simulation.
    }
    \label{fig:jittering}
\end{figure}

Both the particles and the VOF tetrahedra carry accurate Lagrangian momentum information, as compared to the typically more smeared out velocities obtained using semi-Lagrangian advection (see e.g.\ \cite{Stam:1999:SF}) on the Eulerian grid.
Note that we allow VOF tetrahedra to overlap with the Eulerian water.
Thus, we allow for the option to first transfer some momentum from the VOF tetrahedra to the Eulerian grid.
Typically, this increases the turbulence near the boundaries of a moving creature.
This is accomplished by iterating over tetrahedra with water and averaging their momentum with the values on the background Eulerian grid using an artist controllable multiplier.
The result is used to overwrite the value on the Eulerian grid.

Next, the velocities of the Eulerian grid are used to overwrite the momentum value of any tetrahedron which has all four of its nodes inside the water surface representation of the Eulerian grid.
The tetrahedron's volume is also set to be fully saturated with water.
This overwrite operation does not use averaging since the background Eulerian grid has a full-fledged pressure solver that tracks velocities more accurately preserving various effects such as the circulation (discussed in Section~\ref{sec:projection}).
Importantly, cut cell tetrahedra are not overwritten allowing them to more accurately track volume and momentum close to the boundary of the creature.
Higher ranks of tetrahedra could also be allowed to preserve their information if desired, although we did not experiment with this option.

Finally, any particle that lies within a VOF tetrahedron that contains water is deleted, and its volume and momentum are added to that tetrahedron.
This allows particles to freely move through the region of space occupied by the KDSM only being reincorporated into the VOF representation when they impact water regions as defined by the VOF tetrahedra.

\begin{figure}
    \centering
    \includegraphics[width=0.475\textwidth,draft=\mydraft]{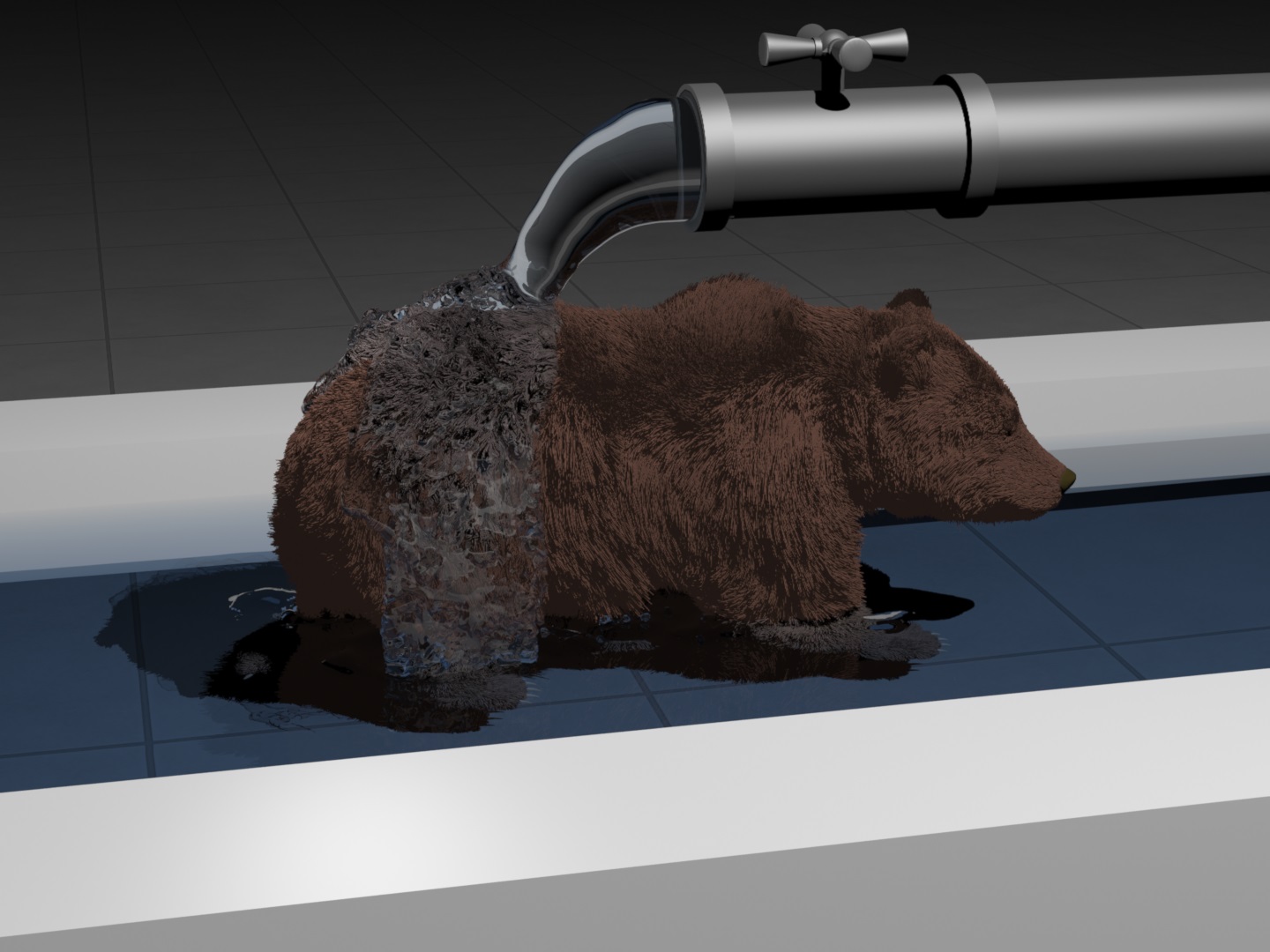}
    \caption{
        Our anisotropic porosity model is implemented to influence the VOF method on the KDSM accounting for both limited volume fraction and drag/adhesion yielding visually compelling results.
    }
    \label{fig:porosity}
\end{figure}

The volume conservation step starts out with the method proposed in Section~\ref{sec:projection}, i.e.\ smear, pushout, and velocity correction, in order to create an adequate velocity for the VOF tetrahedra on the KDSM.
Then, particles are reincorporated into the background Eulerian grid as Eulerian water when appropriate applying a local momentum force, altering the level set, and adding an expansion force similar to \cite{losasso:2008:two}.
Following the standard PLS projection scheme, the results of the pressure solve are subsequently added to the Cartesian grid velocity in order to obtain a divergence free field.
As a final step, the divergence free Eulerian grid velocities are used to overwrite the momentum in any tetrahedron that has all four of its nodes interior to the Eulerian grid water representation.

\subsection{Particle Generation} \label{sec:particle_generation}

\begin{figure*}[t!]
    \centering
    \hbox{
        \includegraphics[width=0.245\textwidth,draft=\mydraft]{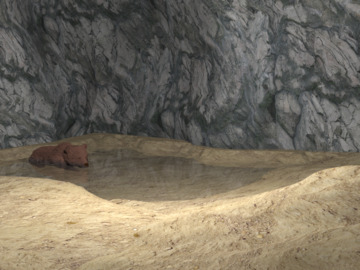}
        \includegraphics[width=0.245\textwidth,draft=\mydraft]{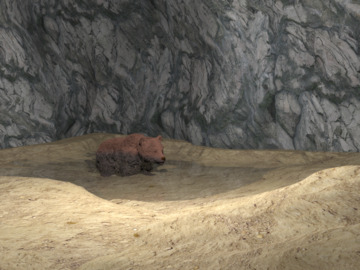}
        \includegraphics[width=0.245\textwidth,draft=\mydraft]{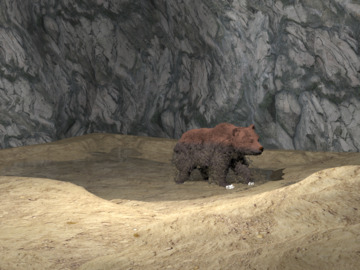}
        \includegraphics[width=0.245\textwidth,draft=\mydraft]{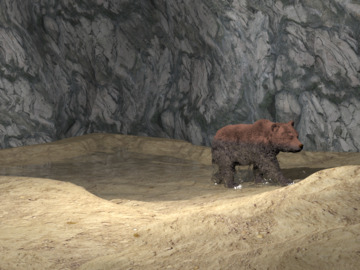}
    }
    \vspace{.017in}
    \hbox{
        \includegraphics[width=0.245\textwidth,draft=\mydraft]{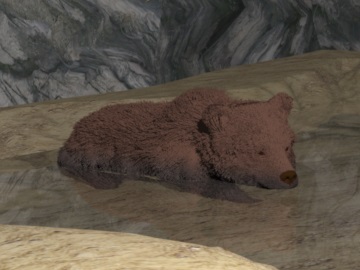}
        \includegraphics[width=0.245\textwidth,draft=\mydraft]{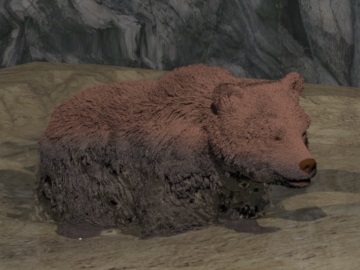}
        \includegraphics[width=0.245\textwidth,draft=\mydraft]{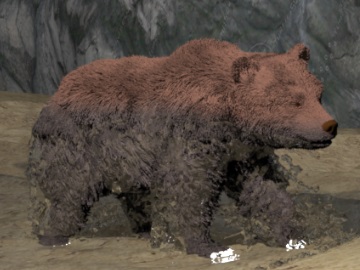}
        \includegraphics[width=0.245\textwidth,draft=\mydraft]{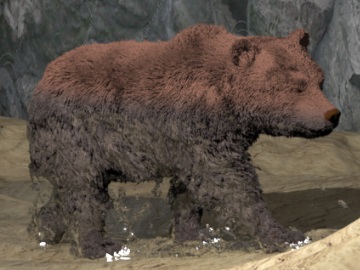}
    }
    \caption{
        A bear walks out of a pool onto land, still carrying and dripping a large amount of water from its fur.
        Our anisotropic porosity model accounts for the correct volume fraction of water in the fur and uses adhesion to pull that water out of the pool with the bear, subsequently slowing dripping the water out of the fur.
        This example emphasizes the efficacy of the adaptivity of the KDSM as well as the ability to preserve volume and avoid disappearing water with our VOF method.
    }
    \label{fig:bear}
\end{figure*}

The automatic generation of particles in visually compelling locations by hybrid particle level set methods has been one of their strengths even predating the PLS method, see \cite{Foster:2001:PAO}.
Thus, we devise a method similar in spirit for our ALE based VOF method on the KDSM.
As discussed in Section \ref{sec:advection}, advection might dictate that water moves off of the KDSM.
This occurs when part of a forward advected tetrahedron lies outside of the KDSM.
This is detected by checking whether or not the point samples of the tetrahedron lie outside of the KDSM.
Each point sample had already been assigned a certain amount of water to transport, so we use that water's volume and momentum to create a particle with appropriate radius and velocity.
Note that we use a standard volume equation for a sphere, $V = 4/3\pi r^3$, to get radius.
Since a straightforward approach leads to noticeable aliasing, we jitter the particle locations by a small amount--we used a fraction ranging from .1 to 1 multiplied by maximum edge length of a tetrahedron for our jitter magnitudes (see Figure \ref{fig:jittering}).
As discussed in Section \ref{sec:projection}, tetrahedra on the exterior boundary of the KDSM may contain excess water that needs to be transported off of the KDSM.
In this scenario, there is no natural advection direction.
Thus, we move the particles across the exterior face of the tetrahedron while also applying appropriate jittering.
Note that when water leaves the KDSM, it always goes through particle phase before rejoining the level set.

\subsection{Rendering} \label{sec:rendering}

As is the case for many of the state-of-the-art Lagrangian methods, rendering smooth surfaces is quite difficult.
Many authors have proposed various strategies, such as applying smoothing kernel on implicit surfaces as in \cite{blinn:1982:sph}, \cite{zhu:2005:sand}, \cite{Adams:2007}, \cite{solen:2007:unified}, \cite{museth:2007:blobtacular}, \cite{williams:2008:reconstruction}, \cite{sin:2009:particle}, \cite{on:2011:sph}, \cite{Bhatacharya:2011:LMS}, \cite{Muller:2011:SSO}, explicitly tracking fluid surfaces as in \cite{brochu:2009:mesh}, \cite{brochu:2010:fronttracking}, and polygonalizing fluid surfaces as in \cite{akinci:2012:efficient}, \cite{akinci:2012:parallel}, \cite{akinci:2013:adaptive}, \cite{wei:2017:reconstruction}.
Since most of this research has been focused on rendering particles as opposed to triangles, we do not use a marching tetrahedra approach as in \cite{doi:1991:marchingtet}, \cite{muller:1997:marchingtet} to render our VOF representation.
Instead, we convert the VOF tetrahedra water into particles and render them along with the other particles.
This is done by creating point samples per tetrahedron based on the quadrature formula (without jittering).
Then, we attract the particles that are near the level set representing the water surface towards that level set in order to flatten out bumps created by the cut cell tetrahedra near the boundary of the level set.
Finally, we use \cite{Yu:2010:RSP} (see also a slight variant of this method \cite{Yu:2013:RSP}) to create an implicit surface from the particle data, and merge this implicit surface with the Eulerian grid level set to obtain the final water surface that we render.
The resulting level set still has some bumps due to the limitations of the anisotropic kernel, so we additionally smooth the normals during rendering.
We stress the fact that we only render the final merged level set on a high resolution grid, and do not directly render the particles.

\section{Hair-Water Interaction} \label{sec:hairwater_interaction}

\begin{figure}[b!]
    \centering
    \includegraphics[width=0.475\textwidth,draft=\mydraft]{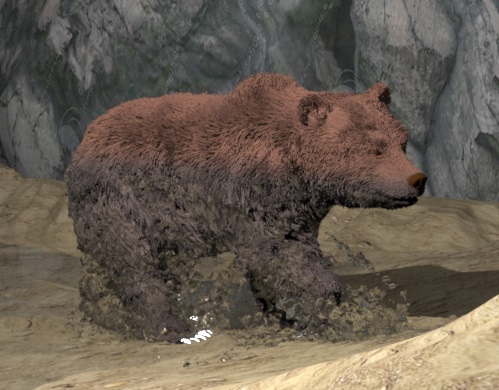}
    \caption{
        A close-up of the bear example, showing water sticking to fur and splashes generated from our VOF method.
    }
    \label{fig:bear_closeup}
\end{figure}

We embed hair particles in the KDSM and treat the hair using the KDSM as in \cite{lee:2018:kdsm_anon} (we also refer the interested reader to \cite{rungjiratananon:2012:wetting}, \cite{lin:2014:coupling}, \cite{Fei:2017:liquidhair}, \cite{Fei:2018:liquidcloth} for more discussion on hair-water interaction).
Our hair-water approach is volumetric in nature, rasterizing multitude of hair representation into KDSM as opposed to \cite{Fei:2017:liquidhair}, where they focus on a reduced model for individual hair strands.
As a result, our method handles hair-water interaction with 540k hairs as opposed to 5k and 30k as given in \cite{rungjiratananon:2012:wetting} and \cite{Fei:2017:liquidhair}, respectively.
For each tetrahedron containing hair we precompute the volume fraction occupied by the hair and reduce the water that this tetrahedron may contain at saturation by this amount.
This gives a very accurate representation of the porosity.
We also compute the average direction of the hair strands in each tetrahedron, so that we may treat the porosity anisotropically.
Essentially, more drag is applied orthogonal to the average direction of the hair strands.
See Figures \ref{fig:porosity}, \ref{fig:bear}, \ref{fig:bear_closeup}.

\section{Results and Discussion}


\begin{figure}
    \centering
    \hbox{
        \includegraphics[width=0.156\textwidth,draft=\mydraft]{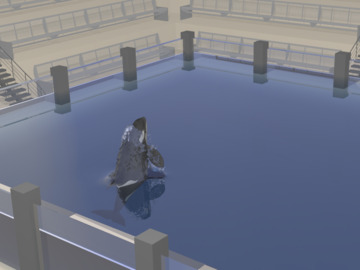}
        \includegraphics[width=0.156\textwidth,draft=\mydraft]{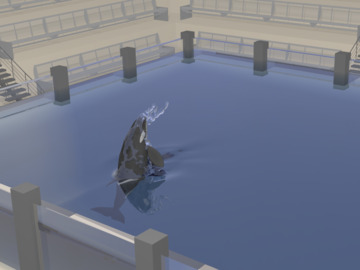}
        \includegraphics[width=0.156\textwidth,draft=\mydraft]{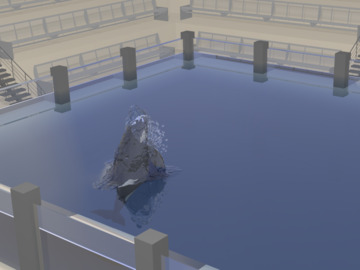}
    }
    \vspace{.01in}
    \hbox{
        \includegraphics[width=0.156\textwidth,draft=\mydraft]{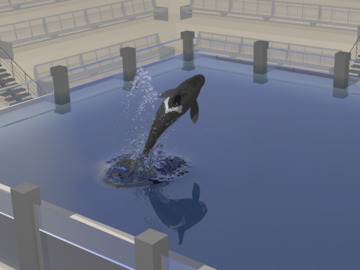}
        \includegraphics[width=0.156\textwidth,draft=\mydraft]{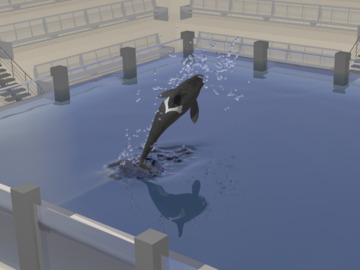}
        \includegraphics[width=0.156\textwidth,draft=\mydraft]{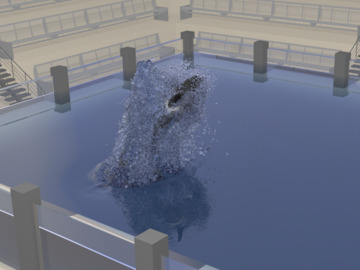}
    }
    \vspace{.01in}
    \hbox{
        \includegraphics[width=0.156\textwidth,draft=\mydraft]{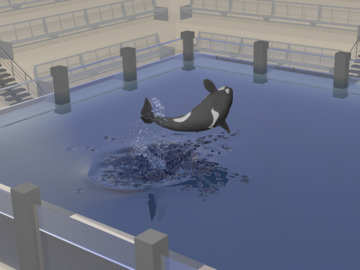}
        \includegraphics[width=0.156\textwidth,draft=\mydraft]{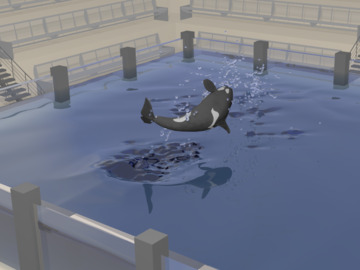}
        \includegraphics[width=0.156\textwidth,draft=\mydraft]{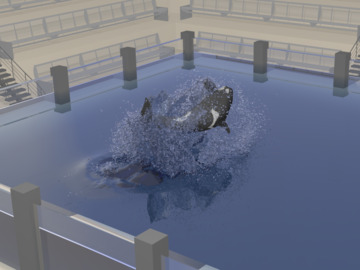}
    }
  \caption{(Left) Whale breaching with the PLS method on high resolution grid. The whale pulls very little water along with it. (Middle) FLIP method, which also produces similar amount of sprays. (Right) Our method pulls more water into the air with the whale, producing interesting effects such as sheets and sprays.}
  \label{fig:whale_compare}
\end{figure}


\begin{figure}[b]
    \centering
    \hbox{
        \includegraphics[width=0.235\textwidth,draft=\mydraft]{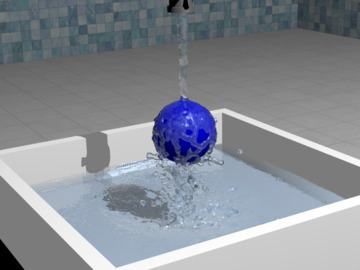}
        \includegraphics[width=0.235\textwidth,draft=\mydraft]{figures/ball_top_vof.jpg}
    }
    \caption{
        (Left) FLIP method on our ball example. (Right) Our method.
    }
    \label{fig:ball_flip}
\end{figure}

We ran our examples on a machine with a 3.06GHz CPU (12 cores) and 96GB RAM.
KDSM generation for the whale and bear examples took 10 minutes per frame, and each frame is temporally independent so we ran them in parallel.
The ball examples (Figures \ref{fig:adhesion_top} and \ref{fig:adhesion_all}) took 1 minute and 2 seconds per frame to run with a 100x100x100 Eulerian grid, 5.6 million KDSM elements, and .9 million KDSM particles.
The bear pour example (Figure \ref{fig:porosity}) took 7 minutes and 3 seconds per frame with a 200x200x400 grid, 8.2 million KDSM elements, and 1.4 million KDSM particles.
The bear walk example (Figure \ref{fig:bear}) took 4.5 minutes per frame with a 100x200x200 grid, 8.2 million KDSM elements, and 1.4 million KDSM particles.
The whale example took 20 minutes per frame with a 200x300x200 grid, 8.5 million KDSM elements, and 1.4 million KDSM particles whereas the PLS method-only example took 29 minutes per frame using a 350x525x350 grid.
We note the visual differences as a comparison in Figure \ref{fig:whale_compare} with FLIP method with a 200x300x200 grid, as well as Figure \ref{fig:ball_flip}.
If we run the PLS method for the whale example at an even higher resolution, we can eventually achieve higher quality results by carrying more water volume with the whale, but this would  require significant time investment.
We used Neumann boundary condition for solid boundaries for PLS method.
For all our examples, we generated 5 to 35 samples per tetrahedron based on the quadrature formula.


There are fundamental limitations of ALE based methods especially regarding meshing problems, so we implement a couple of simple remedies below to fix the occasional degeneracy in order to run all of our examples robustly.
Note that the animated creature can move in a way that inverts its elements or prevents volume preservation of the surrounding space unless the artist is very careful--most of issues appear near joints and are worsened by linear blend skinning.
We only need to iterate a couple of times in the preprocessing stage to resolve most of these issues, and we disable any remaining degenerate elements (inverted or collapsed) so that they cannot participate in the VOF solver.
Thus, whenever a sample point falls in degenerate elements, particles will be formed instead of the located element receiving water.
While one could better prevent element inversion by using FEM or quasistatics, in practice our simple mass spring model was sufficient.
Rarely when we cannot properly advect or enforce incompressibility during the simulation because a VOF tetrahedron is completely surrounded by the solid due to an extreme creature deformation, we simply keep the water in that tetrahedron in order to exactly conserve volume until the issue is resolved as the surrounding solid opens up.

Our method fully conserves volume in the KDSM, although floating point drift causes small volume error throughout the simulation.
We measured the volume error per frame for ball example in top right of Figure \ref{fig:adhesion_top} for 400 frames.
The average volume error per frame was $0.00089\%$, and the maximum error was $0.00189\%$.

Occasionally water stacking along boundaries can occur when VOF tetrahedra are in contact with solids.
This is due to our VOF volume conservation step distributing excess fluid and its momentum to neighboring tetrahedra, and this issue can be resolved either by increasing the resolution of the Eulerian grid to allow Eulerian fluid to contact the solid and using its full-fledged pressure solver as in Section \ref{sec:partitioned_coupling} or by using a standard Poisson solver as discussed in Section \ref{sec:projection}.


\begin{figure}[b]
    \centering
    \includegraphics[width=0.475\textwidth,draft=\mydraft]{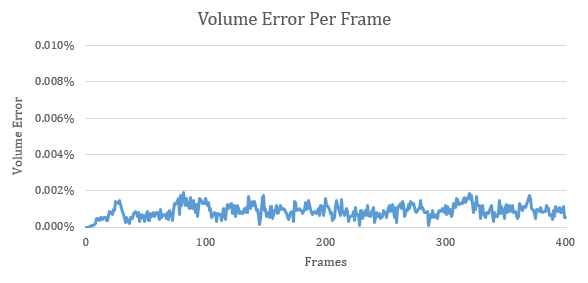}
    \caption{
        Volume error for example where a thin stream of water hits a ball (see top right of Figure \ref{fig:adhesion_top}).
    }
    \label{fig:plot}
\end{figure}

As future work, one could implement a different solver such as \cite{jiang:2015:apic}, \cite{Ihmsen:2012:USF}, or \cite{Chentanez:2015} to simulate fluid in the background grid or in the KDSM, and the adaptivity of our method will improve the accuracy of chosen method.
In order to generalize our method to a pure FLIP/PIC/APIC variant, given that we already have a solver for the background Eulerian grid, the data transfer function would need to be rewritten in order to refer to the KDSM when the particle is inside of the KDSM, and the interpolation scheme would need to be modified to use barycentric weights for tetrahedron.
Then, \cite{ando:2013:tetflip} could be used to handle non-advection steps.
Thus, FLIP/PIC/APIC variant can benefit from the dense KDSM mesh instead of using the coarse background Eulerian grid when the method transfers data from particles to the KDSM.
We emphasize the technical insight that the coarse background grid captures a low frequency fluid surface whereas around the creature with high frequency boundaries we use the dense KDSM mesh to capture high frequency fluid motion.
We chose the PLS method because it generates a very smooth surface, which is suitable for background motion, whereas our ALE based VOF method is more geared towards capturing detailed fluid motion by preserving volume to compensate for the PLS method's limitation.
Additionally, one can subdivide on-the-fly if adaptive remeshing based on the fluid motion is desired.

\section{Conclusions}

We proposed a new fluid simulation framework for character-water and hair-water interaction using our novel volume conserving VOF method based on an adaptive tetrahedral mesh from the KDSM, which moves with the creature.
We prebake the adaptivity of the ALE mesh, separating the nontrivial remeshing issue from the simulation phase and improving the robustness of our ALE based VOF method; we further preprocess auxiliary data wherever possible in order to make the simulation efficient and streamlined.
A coarse background Eulerian grid and our fine ALE mesh are two way coupled using a partitioned approach which is fast, efficient, and straightforward to implement.
We use our volume conserving VOF method only on the KDSM near the creature while using a standard PLS method on the background Eulerian grid.
We robustly implement interesting effects such as adhesion and anisotropic porosity.
We demonstrated how the coarse background Eulerian grid captures the bulk behavior of the water, while our VOF method captures detailed water effects near the creature and the particles capture the spray---all of which make important contributions to the final result.

\section*{Acknowledgements}
The authors would like to acknowledge Ed Quigley for helping us with writing and compositing videos, and Winnie Lin for voice acting.
Research supported in part by ONR N00014-13-1-0346, ONR N00014-17-1-2174, ARL AHPCRC W911NF-07-0027, and NSF CNS1409847.
M.L.\ was supported by Samsung Scholarship.
D.H.\ was supported by NDSEGF.
Computing resources were provided in part by ONR N00014-05-1-0479.

\bibliographystyle{IEEEtran}
\vspace{-.1in}
\bibliography{references}

 \providecommand{\url}[1]{#1}
\begin{thebibliography}{10}
\providecommand{\url}[1]{#1}
\csname url@samestyle\endcsname
\providecommand{\newblock}{\relax}
\providecommand{\bibinfo}[2]{#2}
\providecommand{\BIBentrySTDinterwordspacing}{\spaceskip=0pt\relax}
\providecommand{\BIBentryALTinterwordstretchfactor}{4}
\providecommand{\BIBentryALTinterwordspacing}{\spaceskip=\fontdimen2\font plus
\BIBentryALTinterwordstretchfactor\fontdimen3\font minus
  \fontdimen4\font\relax}
\providecommand{\BIBforeignlanguage}[2]{{%
\expandafter\ifx\csname l@#1\endcsname\relax
\typeout{** WARNING: IEEEtran.bst: No hyphenation pattern has been}%
\typeout{** loaded for the language `#1'. Using the pattern for}%
\typeout{** the default language instead.}%
\else
\language=\csname l@#1\endcsname
\fi
#2}}
\providecommand{\BIBdecl}{\relax}
\BIBdecl

\bibitem{Sussman:2003:ALS}
M.~Sussman, A.~Almgren, J.~Bell, P.~Colella, L.~Howell, and M.~Welcome, ``An
  adaptive level set approach for incompressible two-phase flows,'' \emph{J.
  Comput. Phys.}, vol. 148, pp. 81--124, 1999.

\bibitem{losasso:2004:octree}
F.~Losasso, F.~Gibou, and R.~Fedkiw, ``Simulating water and smoke with an
  octree data structure,'' \emph{ACM Trans. Graph. (SIGGRAPH Proc.)}, vol.~23,
  pp. 457--462, 2004.

\bibitem{Losasso:2005:adaptive}
F.~Losasso, R.~Fedkiw, and S.~Osher, ``Spatially adaptive techniques for level
  set methods and incompressible flow,'' \emph{Computers and Fluids}, vol.~35,
  pp. 995--1010, 2006.

\bibitem{aan:2011:power}
M.~Aanjaneya, M.~Gao, H.~Liu, C.~Batty, and E.~Sifakis, ``Power diagrams and
  sparse paged grids for high resolution adaptive liquids,'' \emph{ACM TOG},
  vol.~36, 2017.

\bibitem{Chentanez:2007:LSL}
N.~Chentanez, B.~E. Feldman, F.~Labelle, J.~F. O'Brien, and J.~R. Shewchuk,
  ``Liquid simulation on lattice-based tetrahedral meshes,'' in \emph{ACM
  SIGGRAPH/Eurographics Symp. on Comput. Anim.}, 2007, pp. 219--228.

\bibitem{Batty:2010:tetrahedral}
C.~Batty, S.~Xenos, and B.~Houston, ``Tetrahedral embedded boundary methods for
  accurate and flexible adaptive fluids,'' in \emph{Comput. Graph. Forum
  (Eurographics Proc.)}, vol.~29, no.~2, 2010, pp. 695--704.

\bibitem{ando:2013:tetflip}
R.~Ando, N.~Th\"urey, and C.~Wojtan, ``Highly adaptive liquid simulations on
  tetrahedral meshes,'' \emph{ACM Trans. Graph. (Proc. SIGGRAPH 2013)}, July
  2013.

\bibitem{english:2012:chimera}
\BIBentryALTinterwordspacing
R.~English, L.~Qiu, Y.~Yu, and R.~Fedkiw, ``An adaptive discretization of
  incompressible flow using a multitude of moving {C}artesian grids,'' \emph{J.
  Comput. Phys.}, vol. 254, no.~0, pp. 107 -- 154, 2013. [Online]. Available:
  \url{http://www.sciencedirect.com/science/article/pii/S0021999113005226}
\BIBentrySTDinterwordspacing

\bibitem{english:2013:chimera_water}
------, ``Chimera grids for water simulation,'' in \emph{SCA '13: Proceedings
  of the 2013 ACM SIGGRAPH/Eurographics symposium on Computer animation}, 2013.

\bibitem{lee:2018:kdsm_anon}
M.~Lee, D.~Hyde, M.~Bao, and R.~Fedkiw, ``A skinned tetrahedral mesh for hair
  animation and hair-water interaction,'' \emph{IEEE TVCG}, 2018.

\bibitem{mihalef:2004:breaking}
V.~Mihalef, D.~Metaxas, and M.~Sussman, ``Animation and control of breaking
  waves,'' in \emph{Proc. of the 2004 ACM SIGGRAPH/Eurographics Symp. on
  Comput. Anim.}, 2004, pp. 315--324.

\bibitem{mihalef:2006:boiling}
V.~Mihalef, B.~Unlusu, D.~Metaxas, M.~Sussman, and M.~Hussaini, ``Physics based
  boiling simulation,'' in \emph{SCA '06: Proc. of the 2006 ACM
  SIGGRAPH/Eurographics Symp. on Comput. Anim.}, 2006, pp. 317--324.

\bibitem{Hirt_Nichols:1981:VOF}
C.~Hirt and B.~Nichols, ``Volume of fluid ({VOF}) method for the dynamics of
  free boundaries,'' \emph{J. Comput. Phys.}, vol.~39, pp. 201--225, 1981.

\bibitem{brackbill:1992:VOF}
J.~U. Brackbill, D.~B. Kothe, and C.~Zemach, ``A continuum method for modelling
  surface tension,'' \emph{J. Comput. Phys.}, vol. 100, pp. 335--353, 1992.

\bibitem{rider:1998:volumetracking}
W.~J. Rider and D.~B. Kothe, ``Reconstructing volume tracking,'' \emph{J.
  Comput. Phys.}, vol. 141, pp. 112--152, 1998.

\bibitem{Sussman:2000:CLSVOF}
M.~Sussman and E.~Puckett, ``A coupled level set and volume-of-fluid method for
  computing {3D} and axisymmetric incompressible two-phase flows,'' \emph{J.
  Comput. Phys.}, vol. 162, pp. 301--337, 2000.

\bibitem{zhang:2016:resolving}
X.~Zhang, M.~Li, and R.~Bridson, ``Resolving fluid boundary layers with
  particle strength exchange and weak adaptivity,'' \emph{ACM Trans. Graph.},
  vol.~35, no.~4, 2016.

\bibitem{omar:2017:positive}
O.~Zarifi and C.~Batty, ``A positive-definite cut-cell method for strong
  two-way coupling between fluids and deformable bodies,'' in \emph{Proceedings
  of the 2016 ACM SIGGRAPH/Eurographics Symposium on Computer Animation}, ser.
  SCA '17, 2017.

\bibitem{Akinci:2013:VST}
\BIBentryALTinterwordspacing
N.~Akinci, G.~Akinci, and M.~Teschner, ``Versatile surface tension and adhesion
  for sph fluids,'' \emph{ACM Trans. Graph.}, vol.~32, no.~6, pp. 182:1--182:8,
  Nov. 2013. [Online]. Available:
  \url{http://doi.acm.org/10.1145/2508363.2508395}
\BIBentrySTDinterwordspacing

\bibitem{feldman:2005:hybrid}
B.~Feldman, J.~O'Brien, and B.~Klingner, ``Animating gases with hybrid
  meshes,'' \emph{ACM Trans. Graph. (SIGGRAPH Proc.)}, vol.~24, no.~3, pp.
  904--909, 2005.

\bibitem{feldman:2005:deforming}
B.~Feldman, J.~O'Brien, B.~Klingner, and T.~Goktekin, ``Fluids in deforming
  meshes,'' in \emph{{Proc. of the ACM SIGGRAPH/Eurographics Symp. on Comput.
  Anim.}}, 2005, pp. 255--259.

\bibitem{Klingner:2006:FAD}
B.~M. Klingner, B.~E. Feldman, N.~Chentanez, and J.~F. O'Brien, ``Fluid
  animation with dynamic meshes,'' \emph{ACM Trans. Graph. (SIGGRAPH Proc.)},
  vol.~25, no.~3, pp. 820--825, 2006.

\bibitem{Enright:2002:PLS}
D.~Enright, R.~Fedkiw, J.~Ferziger, and I.~Mitchell, ``A hybrid particle level
  set method for improved interface capturing,'' \emph{J. Comput. Phys.}, vol.
  183, pp. 83--116, 2002.

\bibitem{Enright:2002:AAR}
D.~Enright, S.~Marschner, and R.~Fedkiw, ``Animation and rendering of complex
  water surfaces,'' \emph{ACM Trans. Graph. (SIGGRAPH Proc.)}, vol.~21, no.~3,
  pp. 736--744, 2002.

\bibitem{losasso:2008:two}
F.~Losasso, J.~O.~Talton, K.~Nipun, and R.~Fedkiw, ``Two-way coupled sph and
  particle level set fluid simulation,'' \emph{IEEE TVCG}, vol.~14, pp.
  797--804, July 2008.

\bibitem{Foster:1996:RAO}
N.~Foster and D.~Metaxas, ``Realistic animation of liquids,'' \emph{Graph.
  Models and Image Processing}, vol.~58, pp. 471--483, 1996.

\bibitem{Ferstl:2016:NBF}
\BIBentryALTinterwordspacing
F.~Ferstl, R.~Ando, C.~Wojtan, R.~Westermann, and N.~Thuerey, ``Narrow band
  flip for liquid simulations,'' \emph{Comput. Graph. Forum}, vol.~35, no.~2,
  pp. 225--232, May 2016. [Online]. Available:
  \url{https://doi.org/10.1111/cgf.12825}
\BIBentrySTDinterwordspacing

\bibitem{molino:2003:meshing}
N.~Molino, R.~Bridson, J.~Teran, and R.~Fedkiw, ``A crystalline, red green
  strategy for meshing highly deformable objects with tetrahedra,'' in
  \emph{12th Int. Mesh. Roundtable}, 2003, pp. 103--114.

\bibitem{ali-hamadi:2013:anatomy}
D.~Ali-Hamadi, T.~Liu, B.~Gilles, L.~Kavan, F.~Faure, O.~Palombi, and M.-P.
  Cani, ``Anatomy transfer,'' in \emph{ACM SIGGRAPH Asia 2013 papers}, ser.
  SIGGRAPH ASIA '13, 2013, pp. 188:1--188:8.

\bibitem{cong:2015:fully}
M.~Cong, M.~Bao, J.~L. E, K.~S. Bhat, and R.~Fedkiw, ``Fully automatic
  generation of anatomical face simulation models,'' in \emph{Proc. of the 14th
  ACM SIGGRAPH / Eurographics Symp. on Comput. Anim.}, 2015, pp. 175--183.

\bibitem{sifakis:2007:hybridsolids}
E.~Sifakis, T.~Shinar, G.~Irving, and R.~Fedkiw, ``Hybrid simulation of
  deformable solids,'' in \emph{Proc. of ACM SIGGRAPH/Eurographics Symp. on
  Comput. Anim.}, 2007, pp. 81--90.

\bibitem{selle:2008:hair}
A.~Selle, M.~Lentine, and R.~Fedkiw, ``A mass spring model for hair
  simulation,'' \emph{ACM Trans. Graph. (SIGGRAPH Proc.)}, vol.~27, no.~3, pp.
  64.1--64.11, Aug. 2008.

\bibitem{muller:2012:fast}
M.~M{\"u}ller, T.-Y. Kim, and N.~Chentanez, ``Fast simulation of inextensible
  hair and fur,'' \emph{VRIPHYS}, vol.~12, pp. 39--44, 2012.

\bibitem{sanchez:2015:real}
R.~S{\'a}nchez-Banderas, H.~Barreiro, I.~Garc{\'\i}a-Fern{\'a}ndez, and
  M.~P{\'e}rez, ``Real-time inextensible hair with volume and shape,'' in
  \emph{Congreso Espa{\~n}ol de Inform{\'a}tica Gr{\'a}fica, CEIG'15}, 2015.

\bibitem{bridson:2003:cloth}
R.~Bridson, S.~Marino, and R.~Fedkiw, ``Simulation of clothing with folds and
  wrinkles,'' in \emph{Proc. of the 2003 ACM SIGGRAPH/Eurographics Symp. on
  Comput. Anim.}, 2003, pp. 28--36.

\bibitem{zhang:2009:quadrature}
L.~Zhang, T.~Cui, and H.~Liu, ``A set of symmetric quadrature rules on triangle
  and tetrahedra,'' 2009.

\bibitem{guendelman:2003:rigid}
E.~Guendelman, R.~Bridson, and R.~Fedkiw, ``Nonconvex rigid bodies with
  stacking,'' \emph{ACM TOG}, vol.~22, no.~3, pp. 871--878, 2003.

\bibitem{Lentine:2011:MMCFS}
M.~Lentine, M.~Aanjaneya, and R.~Fedkiw, ``Mass and momentum conservation for
  fluid simulation,'' in \emph{SCA '11: Proceedings of the 2011 ACM
  SIGGRAPH/Eurographics symposium on Computer animation}, 2011, pp. 91--100.

\bibitem{lentine:2012:water}
M.~Lentine, M.~Cong, S.~Patkar, and R.~Fedkiw, ``Simulating free surface flow
  with very large time steps,'' in \emph{ACM SIGGRAPH/Eurographics Symp. on
  Comput. Anim. 2012}, 2012, pp. 107--116.

\bibitem{Zehnder:2018:ASD}
\BIBentryALTinterwordspacing
J.~Zehnder, R.~Narain, and B.~Thomaszewski, ``An advection-reflection solver
  for detail-preserving fluid simulation,'' \emph{ACM Trans. Graph.}, vol.~37,
  no.~4, pp. 85:1--85:8, Jul. 2018. [Online]. Available:
  \url{http://doi.acm.org/10.1145/3197517.3201324}
\BIBentrySTDinterwordspacing

\bibitem{chorin:1968:NNS}
A.~Chorin, ``{N}umerical solution of the {N}avier-{S}tokes {E}quations,''
  \emph{Math. Comput.}, vol.~22, no.~1, pp. 745--762, 1968.

\bibitem{chorin:1967:ANM}
------, ``A numerical method for solving incompressible viscous flow
  problems,'' \emph{J. Comput. Phys.}, vol.~2, no.~1, pp. 12--26, 1967.

\bibitem{zhu:2014:codimensional}
B.~Zhu, E.~Quigley, M.~Cong, J.~Solomon, and R.~Fedkiw, ``Codimensional surface
  tension flow on simplicial complexes,'' \emph{ACM Trans. Graph. (SIGGRAPH
  Proc.)}, vol.~33, no.~4, pp. 111:1--111:11, 2014.

\bibitem{Stam:1999:SF}
J.~Stam, ``Stable fluids,'' in \emph{Proc. of SIGGRAPH 99}, 1999, pp. 121--128.

\bibitem{Foster:2001:PAO}
N.~Foster and R.~Fedkiw, ``Practical animation of liquids,'' in \emph{Proc. of
  ACM SIGGRAPH 2001}, 2001, pp. 23--30.

\bibitem{blinn:1982:sph}
J.~Blinn, ``A generalization of algebraic surface drawing,'' \emph{ACM Trans.
  Graph. (Proc. SIGGRAPH 1982)}, 1982.

\bibitem{zhu:2005:sand}
Y.~Zhu and R.~Bridson, ``Animating sand as a fluid,'' \emph{ACM Trans. Graph.
  (SIGGRAPH Proc.)}, vol.~24, no.~3, pp. 965--972, 2005.

\bibitem{Adams:2007}
B.~Adams, M.~Pauly, R.~Keiser, and L.~J. Guibas, ``Adaptively sampled particle
  fluids,'' \emph{ACM Trans. Graph. (SIGGRAPH Proc.)}, vol.~26, no.~3, 2007.

\bibitem{solen:2007:unified}
B.~Solenthaler, J.~Schl{\"u}fli, and R.~Pajarola, ``A unified particle model
  for fluid-solid interactions,'' \emph{Computer Animation and Virtual Worlds},
  2007.

\bibitem{museth:2007:blobtacular}
K.~Museth, M.~Clive, and N.~B. Zafar, ``Blobtacular: Surfacing particle systems
  in "pirates of the caribbean 3",'' \emph{SIGGRAPH Sketches}, 2007.

\bibitem{williams:2008:reconstruction}
B.~W. Williams, ``Fluid surface reconstruction from particles,'' Ph.D.
  dissertation, The University of British Columbia, 2008.

\bibitem{sin:2009:particle}
F.~Sin, A.~W. Bargteil, and J.~K. Hodgins, ``A point-based method for animating
  incompressible flow,'' in \emph{Proc. of the 2009 ACM SIGGRAPH/Eurographics
  Symp. on Comput. Anim.}, 2009, pp. 247--255.

\bibitem{on:2011:sph}
J.~Onderik, M.~Chladek, and R.~Durikovic, ``Sph with small scale details and
  improved surface reconstruction,'' \emph{SCCG}, 2011.

\bibitem{Bhatacharya:2011:LMS}
H.~Bhatacharya, Y.~Gao, and A.~Bargteil, ``A level-set method for skinning
  animated particle data,'' in \emph{Proceedings of the 2011 ACM
  SIGGRAPH/Eurographics Symposium on Computer Animation}, ser. SCA '11, 2011.

\bibitem{Muller:2011:SSO}
M.~M\"{u}ller and N.~Chentanez, ``Solid simulation with oriented particles,''
  \emph{ACM TOG}, vol.~30, no.~4, pp. 92:1--92:10, 2011.

\bibitem{brochu:2009:mesh}
T.~Brochu and R.~Bridson, ``Robust topological operations for dynamic explicit
  surfaces,'' \emph{SIAM Journal on Scientific Computing}, vol.~31, no.~4, pp.
  2472--2493, 2009.

\bibitem{brochu:2010:fronttracking}
T.~Brochu, C.~Batty, and R.~Bridson, ``Matching fluid simulation elements to
  surface geometry and topology,'' \emph{ACM Trans. Graph. (SIGGRAPH Proc.)},
  pp. 47:1--47:9, 2010.

\bibitem{akinci:2012:efficient}
G.~Akinci, N.~Akinci, M.~Ihmsen, and M.~Teschner, ``An efficient surface
  reconstruction pipeline for particle-based fluids,'' in \emph{Proceedings of
  Virtual Reality Interactions and Physical Simulations}, 2012.

\bibitem{akinci:2012:parallel}
G.~Akinci, M.~Ihmsen, N.~Akinci, and M.~Teschner, ``Parallel surface
  reconstruction for particle-based fluids,'' \emph{Computer Graphics Forum},
  2012.

\bibitem{akinci:2013:adaptive}
G.~Akinci, N.~Akinci, E.~Oswald, and M.~Teschner, ``Adaptive surface
  reconstruction for sph using 3-level uniform grids,'' \emph{WSCG 2013}, 2013.

\bibitem{wei:2017:reconstruction}
W.~Wu, H.~Li, T.~Su, H.~Liu, and Z.~Lv, ``Gpu-accelerated sph fluids surface
  reconstruction using two-level spatial uniform grids,'' vol.~33, 07 2016.

\bibitem{doi:1991:marchingtet}
A.~Doi and A.~Koide, ``An efficient method of triangulating equi-valued
  surfaces by using tetrahedral cells,'' \emph{IEICE Trans. Inf. {\&} Syst.},
  vol. E74-D, pp. 214--224, 1991.

\bibitem{muller:1997:marchingtet}
H.~M{\"u}ller and M.~Wehle, ``Visualization of implicit surfaces using adaptive
  tetrahedrizations,'' 1997.

\bibitem{Yu:2010:RSP}
J.~Yu and G.~Turk, ``Reconstructing surfaces of particle-based fluids using
  anisotropic kernels,'' in \emph{Proc. of the 2010 ACM SIGGRAPH/Eurographics
  Symp. on Comput. Anim.}, 2010, pp. 217--225.

\bibitem{Yu:2013:RSP}
\BIBentryALTinterwordspacing
------, ``Reconstructing surfaces of particle-based fluids using anisotropic
  kernels,'' \emph{ACM Trans. Graph.}, vol.~32, no.~1, pp. 5:1--5:12, Feb.
  2013. [Online]. Available: \url{http://doi.acm.org/10.1145/2421636.2421641}
\BIBentrySTDinterwordspacing

\bibitem{rungjiratananon:2012:wetting}
W.~Rungjiratananon, Y.~Kanamori, and T.~Nishita, ``Wetting effects in hair
  simulation,'' in \emph{Comput. Graph. Forum}, vol.~31, no.~7.\hskip 1em plus
  0.5em minus 0.4em\relax Wiley Online Library, 2012, pp. 1993--2002.

\bibitem{lin:2014:coupling}
W.~Lin, ``Coupling hair with smoothed particle hydrodynamics fluids,''
  \emph{Proc. of VRIPHYS}, 2014.

\bibitem{Fei:2017:liquidhair}
Y.~Fei, H.~T. Maia, C.~Batty, C.~Zheng, and E.~Grinspun, ``A multi-scale model
  for simulating liquid-hair interactions,'' \emph{ACM Trans. Graph.}, vol.~36,
  no.~4, 2017.

\bibitem{Fei:2018:liquidcloth}
\BIBentryALTinterwordspacing
Y.~R. Fei, C.~Batty, E.~Grinspun, and C.~Zheng, ``A multi-scale model for
  simulating liquid-fabric interactions,'' \emph{ACM Trans. Graph.}, vol.~37,
  no.~4, pp. 51:1--51:16, Aug. 2018. [Online]. Available:
  \url{http://doi.acm.org/10.1145/3197517.3201392}
\BIBentrySTDinterwordspacing

\bibitem{jiang:2015:apic}
C.~Jiang, C.~Schroeder, and J.~Teran, ``An affine particle-in-cell method,''
  \emph{ACM Trans. Graph. (Proc. SIGGRAPH 2015)}, 2015.

\bibitem{Ihmsen:2012:USF}
M.~Ihmsen, N.~Akinci, G.~Akinci, and M.~Teschner, ``Unified spray, foam and air
  bubbles for particle-based fluids,'' \emph{Vis. Comput.}, vol.~28, no. 6-8,
  pp. 669--677, 2012.

\bibitem{Chentanez:2015}
\BIBentryALTinterwordspacing
N.~Chentanez, M.~M\"{u}ller, and T.-Y. Kim, ``Coupling 3d eulerian, heightfield
  and particle methods for interactive simulation of large scale liquid
  phenomena,'' pp. 1--10, 2014. [Online]. Available:
  \url{http://dl.acm.org/citation.cfm?id=2849517.2849519}
\BIBentrySTDinterwordspacing

\end{thebibliography}

\begin{IEEEbiography}[{\includegraphics[width=1in,clip,keepaspectratio]{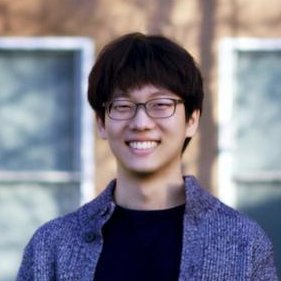}}]{Minjae Lee}
received his Ph.D.\ from Stanford University supported by Samsung Scholarship and MPC-VCC Summer Scholarship. He received his B.S.\ in Computer Science minoring in Art from Carnegie Mellon University in 2012.
His research interests include computer graphics, physically based simulation, computational biochemistry, and games.
\end{IEEEbiography}
\begin{IEEEbiography}{David Hyde}
photography and biography not available at the time of publication.
\end{IEEEbiography}
\begin{IEEEbiography}{Kevin Li}
photography and biography not available at the time of publication.
\end{IEEEbiography}
\begin{IEEEbiography}[{\includegraphics[width=1in,clip,keepaspectratio]{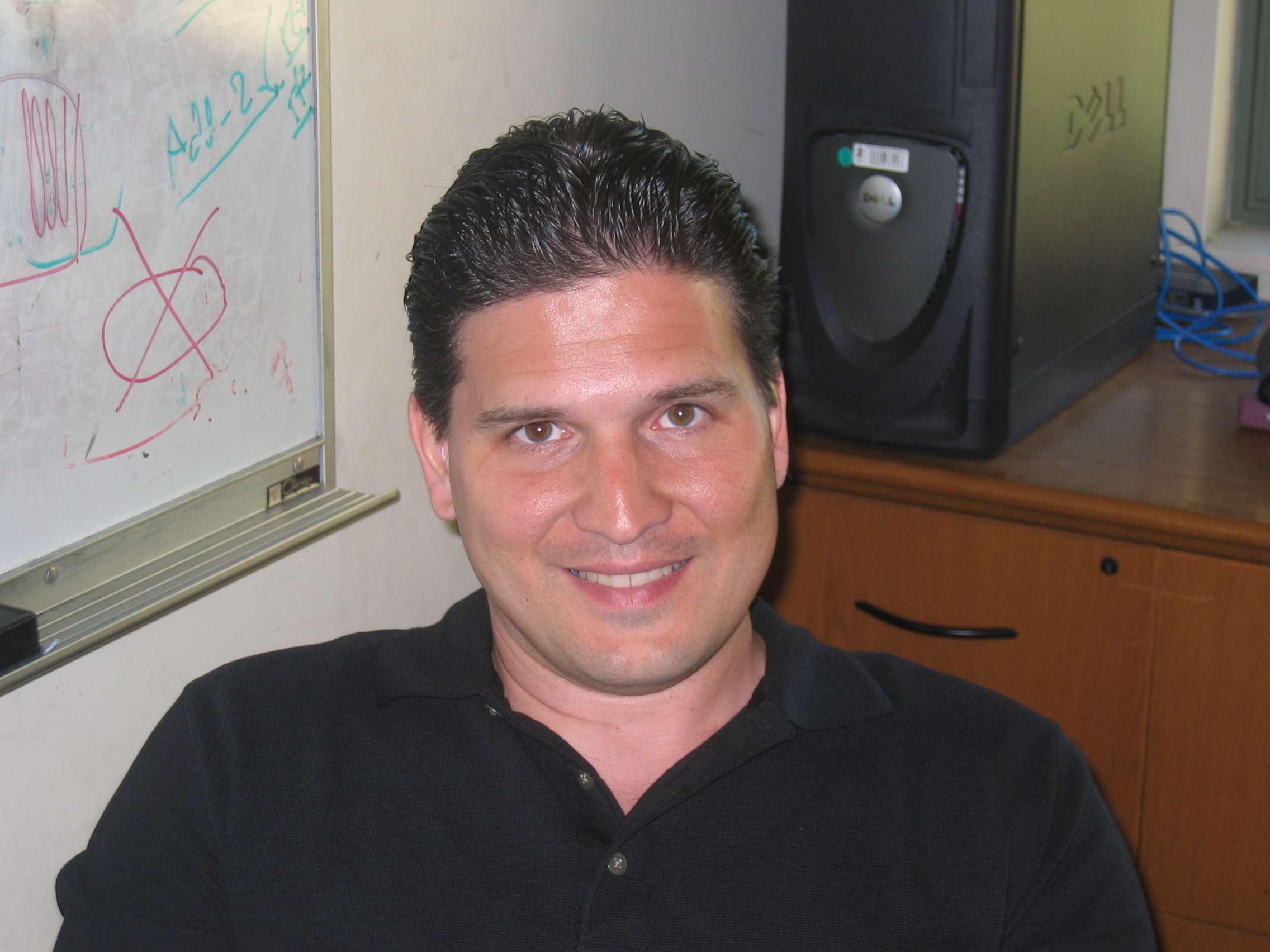}}]{Ronald Fedkiw}
received his Ph.D.\ in Mathematics from UCLA in 1996 and did postdoctoral studies both at UCLA in Mathematics and at Caltech in Aeronautics before joining the Stanford Computer Science Department.
He was awarded an Academy Award from The Academy of Motion Picture Arts and Sciences, the National Academy of Science Award for Initiatives in Research, a Packard Foundation Fellowship, a Presidential Early Career Award for Scientists and Engineers (PECASE), a Sloan Research Fellowship, the ACM Siggraph Significant New Researcher Award, an Office of Naval Research Young Investigator Program Award (ONR YIP), the Okawa Foundation Research Grant, the Robert Bosch Faculty Scholarship, the Robert N. Noyce Family Faculty Scholarship, two distinguished teaching awards, etc.
He has served on the editorial board of the Journal of Computational Physics, Journal of Scientific Computing, SIAM Journal on Imaging Sciences, and Communications in Mathematical Sciences, and he participates in the reviewing process of a number of journals and funding agencies.
He has published over 110 research papers in computational physics, computer graphics, and vision, as well as a book on level set methods.
For the past fifteen years, he has been a consultant with Industrial Light + Magic.
He received screen credits for his work on ``Terminator 3: Rise of the Machines," ``Star Wars: Episode III -- Revenge of the Sith," ``Poseidon," and ``Evan Almighty."
\end{IEEEbiography}

\end{document}